
\magnification=\magstep1
\overfullrule=0pt
\setbox0=\hbox{{\cal W}}
\def\ww{{\cal W}\hskip-\wd0\hskip 0.2 true pt{\cal W}}
\setbox9=\hbox{{\cal S}}
\def\SSWW{{\cal S}\hskip-\wd9\hskip 0.05 true pt{\cal S}}
\def\SW{ \SSWW\ww }
\def\w{{\cal W}}
\def\sw{{\cal SW}}

\def\n{{\cal N}}

\def\lb{\lbrack}
\def\rb{\rbrack}
\def\de{\partial}

\def\q#1{\lb#1\rb}
\def\mn{\medskip\smallskip\noindent}

\def\bn{\bigskip\noindent}

\font\covar=cmssi10 scaled \magstep0
\def\cod{{\hbox{\covar d}}}
\def\coc{{\hbox{\covar C}}}
\def\cochat{\hat \coc }
\def\Cijk{C^{k}_{ij}}
\def\cocijk{\coc^{k}_{ij}}

\def\coctilijk{{\cochat}^{k}_{ij}}
\def\Nuiuj{ {\cal N}(\phi_i,\phi_j) }

\def\cocvivjNuiuj{ \coc^{\Nuiuj}_{{\psi_i}{\psi_j}} }
\def\CvvNuu{ C^{{\cal N}(\phi,\phi)}_{{\psi}{\psi}} }

\def\cowww{C^{\Phi}_{\Phi\Phi}}

\font\extra=cmss10 scaled \magstep0 \font\extras=cmss10 scaled 750

\setbox1 = \hbox{{{\extra R}}}
\setbox2 = \hbox{{{\extra I}}}
\setbox3 = \hbox{{{\extra C}}}

\setbox4=\hbox{{{\extra Z}}}
\setbox5=\hbox{{{\extras Z}}}
\setbox6=\hbox{{{\extras z}}}

\def\z{{{\extras Z}}\hskip-\wd5\hskip 2 true pt{{\extras Z}}}

\def\zed{\hbox{{\extras\z}}}

\def\smno{\smallskip\noindent}
\def\meno{\medskip\noindent}

\def\pano{\par\noindent}
        %
\def\BZT{{\rm Z{\hbox to 3pt{\hss\rm Z}}}}
\def\BZS{{\hbox{\sevenrm Z{\hbox to 2.3pt{\hss\sevenrm Z}}}}}
\def\BZSS{{\hbox{\fiverm Z{\hbox to 1.8pt{\hss\fiverm Z}}}}}

\def\BQT{\,\hbox{\hbox to -2.8pt{\vrule height 6.5pt width .2pt\hss}\rm Q}}
\def\BQS{\,\hbox{\hbox to -2.1pt{\vrule height 4.5pt width .2pt\hss}$
   \scriptstyle\rm Q$}}
\def\BQSS{\,\hbox{\hbox to -1.8pt{\vrule height 3pt width
   .2pt\hss}$\scriptscriptstyle \rm Q$}}

\def\BCT{\,\hbox{\hbox to -3pt{\vrule height 6.5pt width .2pt\hss}\rm C}}
\def\BCS{\,\hbox{\hbox to -2.2pt{\vrule height 4.5pt width .2pt\hss}$
   \scriptstyle\rm C$}}
\def\BCSS{\,\hbox{\hbox to -2pt{\vrule height 3.3pt width
   .2pt\hss}$\scriptscriptstyle \rm C$}}

\def\BHT{{\rm I{\hbox to 5.3pt{\hss\rm H}}}}
\def\BHS{{\hbox{\sevenrm I{\hbox to 4.2pt{\hss\sevenrm H}}}}}
\def\BHSS{{\hbox{\fiverm I{\hbox to 3.5pt{\hss\fiverm H}}}}}

\def\BPT{{\rm I{\hbox to 5pt{\hss\rm P}}}}
\def\BPS{{\hbox{\sevenrm I{\hbox to 4pt{\hss\sevenrm P}}}}}
\def\BPSS{{\hbox{\fiverm I{\hbox to 3pt{\hss\fiverm P}}}}}

\def\BST{\;\hbox{\hbox to -4.5pt{\vrule height 3pt width .2pt\hss}
   \raise 4pt\hbox to -2pt{\vrule height 3pt width .2pt\hss}\rm S}}
\def\BSS{\;\hbox{\hbox to -4.2pt{\vrule height 2.3pt width .2pt\hss}
   \raise 2.5pt\hbox to -4.8pt{\vrule height 2.3pt width .2pt\hss}
   $\scriptstyle\rm S$}}
\def\BSSS{\;\hbox{\hbox to -4.2pt{\vrule height 1.5pt width .2pt\hss}
   \raise 1.8pt\hbox to -4.8pt{\vrule height 1.5pt width .2pt\hss}
   $\scriptscriptstyle\rm S$}}

\def\BFT{{\rm I{\hbox to 5pt{\hss\rm F}}}}
\def\BFS{{\hbox{\sevenrm I{\hbox to 4pt{\hss\sevenrm F}}}}}
\def\BFSS{{\hbox{\fiverm I{\hbox to 3pt{\hss\fiverm F}}}}}

\def\BRT{{\rm I{\hbox to 5.5pt{\hss\rm R}}}}
\def\BRS{{\hbox{\sevenrm I{\hbox to 4.3pt{\hss\sevenrm R}}}}}
\def\BRSS{{\hbox{\fiverm I{\hbox to 3.35pt{\hss\fiverm R}}}}}

\def\BNT{{\rm I{\hbox to 5.5pt{\hss\rm N}}}}
\def\BNS{{\hbox{\sevenrm I{\hbox to 4.3pt{\hss\sevenrm N}}}}}
\def\BNSS{{\hbox{\fiverm I{\hbox to 3.35pt{\hss\fiverm N}}}}}
\def\BN{{\mathchoice{\BNT}{\BNT}{\BNS}{\BNSS}}}
\def\BAT{\hbox{\raise1.8pt\hbox{\sevenrm/}{\hbox to 4pt{\hss\rm A}}}}
\def\BAS{\hbox{\raise1.4pt\hbox{\fiverm/}{\hbox to 3pt{\hss\sevenrm A}}}}
\def\BASS{\hbox{\raise1.4pt\hbox{\fiverm/}{\hbox to 3pt{\hss\sevenrm A}}}}

\def\upin{\hbox{${\scriptstyle\cup}\hbox to -2.7pt{\hss\vrule height
3.5pt}$}}
\def\downin{\hbox{${\scriptstyle\cap}\hbox to -2.7pt{\hss\vrule
height3.5pt}$}}

\def\Sequenz #1,#2,#3{0\longrightarrow #1\longrightarrow #2\longrightarrow
#3 \longrightarrow 0}

\def\teiltnicht{\mathrel{\raise1pt\hbox to .5pt{$\scriptstyle/$\hss}\vert}}
\def\normalspace{}
\def\doublespace{}
\def\bpz{1}
\def\zam{2}
\def\bai{3}
\def\blg{4}
\def\bal{5}
\def\ham{6}
\def\zha{7}
\def\fig{8}
\def\bou{9}
\def\blm{10}
\def\kau{11}
\def\wer{12}
\def\fri{13}
\def\ina{14}
\def\kom{15}
\def\ho1{16}
\def\jos{17}
\def\blu{18}
\def\hor{19}
\def\nam{20}
\def\wes{21}
\def\flo{22}
\def\bo1{23}
\def\cap{24}
\def\ka1{25}
\def\mfl{26}
\def\eho{27}
\def\ber{28}
\def\eic{29}
\def\mus{30}
\def\god{31}
\def\sch{32}
\def\dop{33}
\def\fre{34}
\def\wil{35}
\def\eh1{36}
\def\del{37}
\def\sou{38}
\def\rva{39}
\font\HUGE=cmbx12 scaled \magstep4
\font\Huge=cmbx10 scaled \magstep4
\font\Large=cmr12 scaled \magstep3
\font\MLarge=cmti12 scaled \magstep3
\font\large=cmr17 scaled \magstep0
%
%
\nopagenumbers
\pageno = 0
\centerline{\HUGE Universit\"at Bonn}
\vskip 10pt
\centerline{\Huge Physikalisches Institut}
\vskip 2cm
\centerline{\Large New {\MLarge N}=1 Extended Superconformal}
\vskip 6pt
\centerline{\Large Algebras}
\vskip 6pt
\centerline{\Large with Two and Three Generators}
\vskip 0.8cm
\centerline{(to be published in Int.\ Jour.\ of Mod.\ Phys.\ {\bf A})}
\vskip 0.8cm
\centerline{\large R.\ Blumenhagen, W.\ Eholzer, A.\ Honecker, R.\ H\"ubel}
\vskip 1.5cm
\centerline{\bf Abstract}
\vskip 15pt
\noindent
In this paper we consider extensions of the super Virasoro algebra
by one and two super primary fields. Using a non-explicitly covariant
approach we compute all $\sw$-algebras with one generator of dimension
up to $7$ in addition to the super Virasoro field. In complete analogy to
$\w$-algebras with two generators most results can be classified
using the representation theory of the super Virasoro algebra.
Furthermore, we find that the $\sw({3\over2},{11\over2})$-algebra
can be realized as a subalgebra of $\sw({3\over2},{5\over2})$
at $c = {10\over7}$. We also construct some new $\sw$-algebras
with three generators, namely $\sw({3\over2},{3\over2},{5\over2})$,
$\sw({3\over2},2,2)$ and $\sw({3\over2},2,{5\over2})$.
\vfill
\settabs \+&  \hskip 110mm & \phantom{XXXXXXXXXXX} & \cr
\+ & Post address:                       & BONN-HE-92-02   & \cr
\+ & Nu{\ss}allee 12                     & hep-th/9207072  & \cr
\+ & W-5300 Bonn 1                       & Bonn University & \cr
\+ & Germany                             & January 1992    & \cr
\+ & e-mail: unp039@ibm.rhrz.uni-bonn.de & ISSN-0172-8733  & \cr
\eject
\pageno=1
\footline{\hss\tenrm\folio\hss}
\smno
{\bf 1. Introduction} \meno
Extended conformal and superconformal algebras have gained intensive
interest in theo\-retical physics because they can be considered as
symmetry algebras of conformal quantum field theories (CFT). One
hopes to describe all rational conformal quantum field theories (RCFT) as
minimal models of extended chiral algebras $\q{\bpz}$, so-called
$\w$-algebras.
\pano
The first $\w$-algebras were constructed by Zamolodchikov
in 1985 $\q{\zam}$. He
extended the Virasoro algebra by primary fields up to dimension three.
Two of these $\w$-algebras have the typical property that normal ordered
products appear.
We denote the extension of the Virasoro algebra by primary fields of
conformal dimensions $\delta_1,\ldots,\delta_n$ by
$\w(2,\delta_1,\ldots,\delta_n)$.
In the last years, several methods have been developed
to find new $\w$-algebras. One approach uses the free field construction
based on Lie algebras and
their infinite dimensional extensions, the Kac-Moody algebras
$\q{\bai}\q{\blg}\q{\bal}$.
\smno
In this paper we use a different method, constructing the algebra
explicitly. Using this method a whole bunch of $\w$-algebras has been
constructed
$\q{\ham}\q{\zha}\q{\fig}\q{\bou}\q{\blm}\q{\kau}$.
In $\q{\blm}\q{\kau}$ a constructive algorithm has been presented
which is based on a structural theorem about chiral $SU(1,1)$
invariant algebras $\q{\wer}$.
It has been shown in $\q{\bou}$ that $\w$-algebras with
two generators can only exist
for generic values of the central charge, if the conformal
dimension of the additional primary field is contained in
$\lbrace{1\over
2},1,{3\over2},2,3,4^{\q{\zha}\q{\ham}},6^{\q{\fig}}\rbrace$.
In $\q{\blm}\q{\kau}$ it has been shown that several other
$\w$($2,\delta$)-algebras exist for finitely many values of
the central charge $c$.
The possible $c$ values are essentially determined
by the representation theory of the Virasoro algebra.
\smno
RCFT's describe two dimensional statistical models at second order
phase transitions.
Furthermore there are statistical models which possess superconformal
invariance $\q{\fri}$ at criticality. These models are described by minimal
models of the super Virasoro algebra. Similar to the Virasoro algebra,
the minimal series is bounded from above by $c={3\over 2}$. In order
to get minimal models with higher values of the central charge
one investigates
extensions of the super Virasoro algebra, called super $\w$-algebras.
\smno
In this paper we consider extensions of the super Virasoro
algebra by one and two super primary fields. In general
we denote the extension of the super Virasoro algebra by super primary
fields of dimension
$d_1,\ldots,d_n$ by $\sw({3\over2},d_1,\ldots,d_n)$.
Some $\sw({3\over2},d)$-algebras have already been calculated
in $\q{\ina}\q{\kom}\q{\ho1}\q{\jos}$.
As far as one knows $\sw({3\over2},d)$-algebras exist generically
only for $d \in \lbrace{1\over 2},1,{3\over 2},2\rbrace$.
For $d \ne 2$ these algebras are super Lie algebras.
In $\q{\jos}$ all $\sw({3\over2},d)$-algebras up to $d={7\over 2}$
have been
constructed using the super conformal bootstrap method. Here we extend
this series of algebras up to $d=7$ using the same methods as
presented in $\q{\blm}$.
Superconformal covariance
is implemented by considering Jacobi identities involving the
generator $G(z)$ of pure super transformations.
This algorithm differs from the explicitly covariant one presented in
$\q{\blu}$.
Nevertheless these two algorithms seem to be equivalent.
For $d \ge {5\over 2}$ the $\sw({3\over2},d)$-algebras exist only
for
discrete values of $c$. Most of these values can be classified in
complete analogy to the $\w(2,\delta)$-algebras, essentially using
representation theory  of the super Virasoro algebra.
Furthermore, we find that the
$\sw({3\over2},{11\over2})$-algebra can be realized as a
subalgebra of the $\sw({3\over2},{5\over2})$-algebra
for $c={10\over 7}$. A similar construction has been carried out
in $\q{\hor}$
where it has been shown that $\sw({3\over2},{7\over2})$
is contained
in $\sw({3\over2},{3\over2})$ for $c={7\over5}$.
\smno
This paper is organized as follows: In the next chapter we review
a general theorem about chiral $SU(1,1)$ invariant algebras. In the
third chapter we give the general outline for the construction of
$\sw$-algebras
in the non-explicitly covariant approach. Chapter four contains
our results on $\sw({3\over2},d)$-algebras.
Afterwards we discuss  these results in chapter five.
We continue with our results concerning $\sw({3\over2},d_1,d_2)$-algebras
in chapter six. In chapter seven we draw some conclusions from the
results obtained.
\bn
\leftline{{\bf 2. General theorems about SU(1,1) invariant algebras}}
\mn
This chapter is devoted to a short review of well known results
concerning
algebras of local chiral fields. We will only  give a brief summary;
for proofs as well as for details we refer to $\q{\nam}\q{\blm}$.
\mn
Let $\cal F$ be the algebra of local chiral fields of
a conformal field theory defined on 2-dimensional spacetime (with
compactified space). Because of $SU(1,1)$-invariance $\cal F$ carries
a natural grading by the conformal dimension and is
spanned by the non-derivative (i.\ e.\ quasiprimary) fields
together with their derivatives. In $\cal F$ the operation of
building normal ordered products (NOPs) is defined (see below).
\bn
We define the Fourier decomposition of a left chiral field by $\phi(z)=
\sum_{n-d(\phi)\in\zed} z^{n-d(\phi)}\phi_n\,$. We  call the
Fourier-components
$\phi_n$ the `modes' of $\phi$.
\pano
Denote the vacuum of the theory by $\mid\! v\rangle\,$.
Requiring the regularity of $\phi(z) \mid \!v \rangle\, $ at the origin
implies
$$\phi_n \mid \!v \rangle\, = 0 \ \ \ \ \forall \  n < d(\phi) \eqno({\rm
2.0})$$
It is well known that the modes of the energy momentum tensor
satisfy the Virasoro algebra
$$ \lb L_m,L_n\rb = (n-m)L_{m+n}+{c\over12}(n^3-n)\delta_{n+m,0}
\eqno({\rm 2.1}) $$ with central charge $c$.
A primary field $\phi$ of conformal dimension $d$ is characterized
by the commutator of its modes with the Virasoro algebra (2.1):
$$ \lb L_m,\phi_n\rb = (n - (d-1) m)\phi_{n+m} \eqno({\rm 2.2}) $$
Iff a field $\phi$ satisfies (2.2) for $m \in \{ -1, 0, 1 \} $ this field
is
called `quasiprimary'.
\mn
As was shown by W. Nahm $\q{\nam}\q{\wer}$
locality and invariance under rational conformal transformations
already impose severe restrictions on
the commutator of two quasiprimary chiral fields:
\mn
Let $\lbrace\phi_i\mid i\in I\rbrace$ be
a set of non-derivative fields of integer or half
integer dimensions $d(\phi_i)=d_i$, which together
with their derivatives span $\cal F$.
Then the mode algebra of the Fourier components of
left chiral fields has the form
$$\lb\phi_{i,m},\phi_{j,n}\rb_{\pm}=\sum_{k\in
I}C^{\phi_k}_{\phi_i\phi_j}\,p_{ijk}(m,n)
\phi_{k,m+n} + d_{ij}\,\delta_{n,-m}{n+d_i-1\choose
2d_i-1}\eqno{{\rm(2.3)}}$$
Here the $d_{ij}$ describe the normalization of the two point
functions and the $C^{\phi_k}_{\phi_i\phi_j}$ are the coupling
constants between three   quasiprimary fields.
The $p_{ijk}$ are universal polynomials depending only on the dimensions
of the fields $\phi_i,\phi_j$ and $\phi_k$;
their explicit form is presented in $\q{\blm}$.
\mn
In addition to their Lie bracket structure the algebra $\cal F$ admits
another important operation, namely forming normal ordered
products (NOPs) of chiral fields. Usually the NOP of two
chiral fields $\phi,\chi$ is defined in terms of Fourier
components as follows
$$N(\phi,\chi)_n :=\epsilon_{\phi\chi}
\sum_{k<d(\chi)}\!\phi_{n-k}\chi_k\,+
\sum_{k\geq d(\chi)}\!\chi_k\phi_{n-k} \  \eqno({\rm 2.4}) $$
$\epsilon_{\phi\chi}$ is defined as $-1$ if $\phi$ and $\chi$ are both
fermions and 1 otherwise.
\mn
In this form it occurs in the regular part of the OPE of $\phi$
and $\chi$,
but it is not a non-derivative field, so that e.g.
equation (2.3) cannot be used to gain any information about its
commutator with other fields. For the NOP to be `well behaved'
under $SU(1,1)$-transformations we have to add some counterterms to
$N(\phi,\chi)$ defined in (2.4):
\mn
With the assumptions and notations from
above we define the
normal ordered product of two chiral fields by $\q{\wer}$
$$ \n(\phi_j,\partial^{n}\phi_i) :=
          N(\phi_j,\de^{n}\phi_i)
-  \bigl( \sum_{r=1}^{n}
          \alpha^r_{ij} \de^r N(\phi_j,\de^{n-r}\phi_i)
          + \!\!\sum_{k} \beta^k_{ij}(n)
          C^{\phi_k}_{\phi_i\phi_j} \partial^{\gamma^k_{ij}(n)}\phi_k
   \bigr)
  \eqno( {\rm 2.5} )
$$
Where $\alpha^r_{ij},\beta^k_{ij}(n),\gamma^k_{ij}(n)$ are
polynomials depending only on the conformal dimensions
of the involved fields (for more details see $\q{\blm}$). This definition
yields a quasiprimary field of conformal dimension $ d_i+d_j+n $.
\mn
Since the field content of $\cal F$ is infinite one introduces
the notion of non-composite, `simple' fields.
To be more precise, if a basis for
$\cal F$ can be obtained from a set of fields $\{ \phi_i \}_{i\in J}$
using the
operations given above (forming normal ordered products and derivatives)
we will say that the fields $\{ \phi_i \}_{i\in J}$ generate $\cal F$.
If all fields in $\{ \phi_i \}_{i\in J}$ are quasiprimary and orthogonal
to all normal ordered products
we will call these fields `simple'. A set of simple fields
will be called a simple set.
\mn
The commutators of normal ordered products are
completely determined by the commutators of the simple fields involved.
This means that the whole Lie algebra structure of the mode algebra
of $\cal F$ is already fixed by the commutation relations of the
simple fields. The coupling constants of the simple fields determine all
other coupling constants.
\mn
In general, $\w$-algebras are not Lie algebras since the commutators
do not close
linearly in the fields. If the OPE corresponding to the mode algebra is
required to be associative, the mode algebra has to fulfil all
possible Jacobi identities.
In order to ensure the validity of all Jacobi identities  it is
sufficient to verify that only those involving simple fields (besides $L$)
are satisfied.
Looking at the corresponding OPE one notices that it is sufficient
to show that the factors in front of primary fields vanish.
This leads to equations among the coupling constants connecting three
simple fields.
\bn
\leftline{{\bf 3. General definitions and theorems about $\SW$-algebras  }}
\mn
In this chapter we adapt the results of the conformal case to the
superconformal one. To this end, we recall the definitions of super
fields, super primaries and super Jacobi identities. Furthermore, we
set up
some conventions and establish some simple relations between various
structure constants arising from superconformal covariance.
\mn
The super Virasoro algebra is the extension of the Virasoro algebra
by a primary field $G$ of dimension $3\over2$.
Using the normalization $d_{GG} = {2c\over3}$, the commutation
relations take the following form:
\pano
$$ \eqalign{ \lb L_m,L_n\rb &= (n-m)L_{m+n}+{c\over12}(n^3-n)\delta_{n+m,0}
\cr
             \lb L_m,G_n\rb &= (n-{1\over2}m)G_{m+n} \cr
             {\lb G_m,G_n\rb}_{+} &= 2L_{m+n} +
{c\over3}(m^2-{1\over4})\delta_{m+n,0},
            } \eqno( {\rm 3.1}) $$
which can also be obtained from the theorem of chapter 2.
This algebra is known as the central extension of the algebra
formed by the generators of superconformal transformations in
super space $Z$ consisting of the points $(z,\theta)$, where
$\theta$ is a Grassmannian variable.
The two fields can be composed to the super Virasoro field
${\cal L} = {1\over2}G + \theta L$, defined on $Z$.
\pano
A field $\Phi = \phi + \theta\psi$  is called super primary of dimension
$d = d(\Phi)$, iff $\phi$ and $\psi$ are Virasoro primaries of conformal
dimension $d$ and $d+{1\over2}$ respectively and
$$\eqalign{  {\lb G_n,\phi_m\rb}_{\pm} &= \coc^{\psi}_{G\phi}
p_{{3\over2},d,d+{1\over2}}(n,m)\  \psi_{n+m} = \coc^{\psi}_{G\phi}
\psi_{n+m} \cr
             {\lb G_n,\psi_k\rb}_{\pm} &= \coc^{\phi}_{G\psi}
p_{{3\over2},d+{1\over2},d}(n,k)\  \phi_{n+k} =
{\coc^{\phi}_{G\psi}\over{2d}}(k-(2d-1)n) \phi_{n+k},
          } \eqno( {\rm 3.2} ) $$
with $\phi$ and $\psi$ super partners of each other;
here $\coc^{\psi}_{G\phi}$ and  $\coc^{\phi}_{G\psi}$ are certain
structure constants and we have
inserted the universal polynomials $p_{ijk}$.
In perfect analogy to the quasiprimary case a field $\phi +\theta\psi$
is called super quasiprimary iff
$\phi$ and $\psi$ are Virasoro quasiprimaries and (3.2) holds for
$ n \in \{ {1\over2},-{1\over2}\} $.
This definition is equivalent to the covariant definition of super
primary fields $\q{\wes}$.
\mn
Let  ${\cal A} = \{G,L,\phi_1,\psi_1,...,\phi_n,\psi_n\}$
be a simple set with super primaries $\Phi_i = \phi_i +\theta\psi_i$.
We will call the algebra generated by $\cal A$ a
$\sw$(${3\over2},d(\Phi_1)$,...,$d(\Phi_n)$) and the fields
$\Phi_1,...,\Phi_n$ additional fields.
\mn
To ensure the existence of  $\sw({3\over2},d_1,...,d_n)$
super Jacobi identities have to be checked which are of the general
form:
$$ { \bigl( \epsilon_{\chi_i\chi_k}
     {\lb {\lb \chi_{i,p} , \chi_{j,q} \rb}_{\pm} ,\chi_{k,r} \rb }_{\pm}
     \bigr)
   }_{cycl.} = 0, \eqno{(3.3)} $$
where $\chi_i,\chi_j$ and $\chi_k$ are components ($\phi$ or $\psi$) of
super fields.
In the sequel we will denote the factor in front of the field
$\chi_{l,p+q+r}$ on the left hand side of $(3.3)$
by $(\chi_{i,p},\chi_{j,q},\chi_{k,r},\chi_{l,p+q+r})$.
\pano
The simplest example for $\sw$-algebras is the super Virasoro algebra
$(3.1)$ since it obeys the super Jacobi identities $(3.3)$, which is also
a super Lie algebra.
\mn
The identities $(3.3)$ will lead to relations involving different structure
constants. Some of these identities hold in general, already determined by
the
supersymmetric structure. This illustrates the fact that we are working
with a non-explicitly covariant approach, dealing with components of super
fields rather than with super fields themselves.
\pano
These relations between various structure constants can be derived from
simple Jacobi identities involving the component $G$ of the
super Virasoro field. In order to give some examples we
fix the normalization of the fields first:
$$\eqalign{ \cod_{{\phi_i}{\phi_i}} &= {(-1)^{2d(\phi_i)+1}\over
2d(\phi_i)+1}
                                    {c\over d(\phi_i)}   \cr
            \cod_{{\psi_i}{\psi_i}} &= { c\over d(\psi_i) }
          } \eqno( {\rm 3.4} ) $$
Note that in the covariant approach this leads to the standard
normalization  of super fields $\q{\blu}$.
\mn
Then we can e.g. determine the structure constants $\coc^{\psi}_{G\phi}$
and $\coc^{\phi}_{G\psi}$ in $(3.2)$:
\pano
 $$\coc^{\psi_i}_{G\phi_i} = {1\over{2d_i}} \coc^{\phi_i}_{G\psi_i}
   \ \ \ \ \ {\rm{ and }} \ \ \ \ \ \
   {\bigl( \coc^{\psi_i}_{G\phi_i} \bigr)}^2 = 1 \eqno({\rm 3.5})$$
\pano
The first of these relations between the different structure
constants is a direct consequence of the normalization of the
fields $\phi_i$ and $\psi_i$ and the fact that they are simple fields.
The other relation is easily seen for  $d_i \in \BN$ by the following
argument (the case $d_i \in \BN+{1\over2}$ is treated similarly):
\ Consider the following Jacobi identity:
$$ {\bigl( {\lb {\lb G_r , G_s \rb}_{\pm} ,
\phi_{i,m} \rb }_{\pm}\bigr)}_{cycl.} = 0 $$
Because $(G_r , G_s ,\phi_{i,m},\phi_{i,r+s+m})$
 has to be zero we obtain:
\mn
$$ \coc^{L}_{GG}\coc^{\phi_i}_{L\phi_i}p_{2,{d_i},{d_i}}(r+s,m) - $$
$$ \coc^{\psi_i}_{G\phi_i}\coc^{\phi_i}_{{\psi_i}G} \bigl(
p_{{3\over2},{d_i},{d_i}+{1\over2}}(s,m)p_{{d_i}+{1\over2},{3
\over2},{d_i}}(s+m,r)
+p_{{d_i},{3\over2},{d_i}+{1\over2}}(m,r)p_{{d_i}+{1\over2},{3
\over2},{d_i}}(m+r,s) \bigr)  = 0 $$
\mn
\leftline{{ using ${\coc^{\phi_i}_{{\psi_i}G}} =
                  {2d_i}\coc^{\psi_i}_{G\phi_i}
 ,\coc^{L}_{GG} = 2 $ and  $ \coc^{\phi_i}_{L\phi_i} = d_i$, \   one
obtains: }}
   $${\bigl( \coc^{\psi_i}_{G\phi_i}\bigr)}^2 = 1  $$
\pano
In this chapter we will choose the positive root, so that
$\coc^{\psi_i}_{G\phi_i} = 1$.
\mn
Furthermore, all possible coupling constants between three super
primary fields
$(\phi_i,\psi_i)\times(\phi_j,\psi_j)\rightarrow(\phi_k,\psi_k)$ are
determined by a single coupling constant of the components:
\mn
\leftline{{\  i)   $d_i+d_j+d_k \in \BN + {1\over2} $: }}
$$ \coc^{\phi_k}_{{\psi_i}{\phi_j}} = {2d_k}\cocijk \ \  ,
\ \ \coc^{\phi_k}_{{\phi_i}{\psi_j}} = (-1)^{2d_i+1}{2d_k}\cocijk \ \ ,
\ \ \coc^{\psi_k}_{{\psi_i}{\psi_j}} = (-1)^{2d_i+1}
    (\sigma_{ijk}+{\scriptstyle{1\over2}})\cocijk
 \eqno({\rm3.6a}) $$
\ \ with $ \cocijk = \coc^{\psi_k}_{{\phi_i}{\phi_j}}$
\mn
\leftline{{ ii)   $d_i+d_j+d_k \in \BN $: }}
$$ \coc^{\psi_k}_{{\psi_i}{\phi_j}} = {  {h_{ikj}}\over{2d_k}  }
                                  \coctilijk \ \ ,
\ \ \coc^{\phi_k}_{{\psi_i}{\psi_j}} = (-1)^{2d_i+1}h_{ijk}
                                  \coctilijk \ \ ,
\ \ \coc^{\phi_k}_{{\phi_i}{\psi_j}} = (-1)^{2d_i}
                             {{h_{jki}}\over{2d_k}} \ \coctilijk
 \eqno({\rm3.6b}) $$
\ \ \ \ \ \ \ with $ \coctilijk = \coc^{\phi_k}_{{\phi_i}{\phi_j}}$
\mn
and $ \sigma_{ijk} = d_i+d_j+d_k-1 \ , \ h_{ijk} = d_i+d_j-d_k $
\mn
\leftline{{For a proof consider  the condition }}
$$(G_l ,\phi_{i,m} ,\phi_{j,n}, \psi_{k,l+m+n}) = 0 .$$
This yields an equation in the coupling constants
$$  \coc^{\psi_k}_{{\psi_i}{\phi_j}} \ \ ,
\ \ \coc^{\psi_k}_{{\phi_i}{\psi_j}}  \ \ , \ \
\ \ \coc^{\phi_k}_{{\phi_i}{\phi_j}} . $$
Setting $ l =  {1\over2}\ ,\ m = d_j-d_k \ ,\ n = 1-d_j $ and
        $ l = -{1\over2}\ ,\ m = 1-d_i \ ,\ n = d_i-d_k $  one obtains
two equations. Combining these equations  proofs the first
two relations.
The other relations are obtained by studying
$$ \eqalign{
  ( G_l ,\phi_{j,m} ,\psi_{i,n},\phi_{k,l+m+n}) &= 0 \cr
  ( G_l ,\psi_{i,m} ,\psi_{j,n},\phi_{k,l+m+n}) &= 0 \cr
  ( G_l ,\psi_{j,m} ,\phi_{i,n},\psi_{k,l+m+n}) &= 0.
}$$
\mn
{\bf Remark 3.1:}
If one assumes that a basis of super quasiprimary fields exists
(e.g. in the covariant approach)  the formulae given above
hold also for super quasiprimary fields since
only $l \in \{{1\over2},-{1\over2} \} $  is needed for the proof.
The same formulae can be deduced in the explicitly covariant approach
$\q{\blu}$.
\mn
{\bf Remark 3.2:} Because of $\cocijk =
(-1)^{\lb{d_i}\rb+\lb{d_j}\rb+\lb{d_k+{1\over2}}\rb}\coc^{k}_{ji}$
and  $\coctilijk =
(-1)^{\lb{d_i}\rb+\lb{d_j}\rb+\lb{d_k}\rb}{\cochat}^{k}_{ji}$ one obtains
for $(\phi_i,\psi_i) = (\phi_j,\psi_j)$:
\mn
\leftline{{ \ i)  ${\cochat}^{k}_{ii}\ne 0 \ \ \Rightarrow  \ \
                {d_k}\in 2\BN$}}
\pano
\leftline{{ ii) ${\coc}^{k}_{ii}\ne 0 \ \ \Rightarrow  \ \
                {d_k +{1\over2}}\in 2\BN$}}
\mn\mn
In the explicitly covariant formulation $\q{\blu}$ it is
evident
that in the singular part of the OPE of a super primary field
$\Phi_i$ with $\Phi_j$ no normal ordered products of
these super fields appear.
In the non-explicitly covariant approach this may happen only
if the field $\Nuiuj$ appears in the commutator
$\lb \psi_i,\psi_j \rb_{\pm}$. In this case  either
$\Nuiuj$ must have a vanishing primary projection and
can be replaced by a linear combination of the other fields of
dimension $d_i+d_j$ or
the relevant coupling constant has to vanish.
Denote the primary projection of $\phi$ by $\cal P \phi$.
Then one has
$$  {\cal P} \Nuiuj \ne 0 \ \ \  \Rightarrow \ \ \ \cocvivjNuiuj = 0
\eqno({\rm 3.7})$$
\mn
Since
$$ {\lb \psi_{i,m},\psi_{j,n}\rb}_{\pm} =
   \cocvivjNuiuj\Nuiuj_{m+n} + ...\ \  ,$$
applying $G_{{1\over2}}$ to ${\lb \psi_{i,m},\psi_{j,n}\rb}_{\pm}$ yields
$$ \eqalign{
 \coc^{{\cal N}(\phi_i,\psi_j)}_{G\Nuiuj}{\cocvivjNuiuj}
   &{\cal N}(\phi_i,\psi_j)_{n+m+{1\over2}}+
 \coc^{{\cal N}(\psi_i,\phi_j)}_{G\Nuiuj}{\cocvivjNuiuj}
   {\cal N}(\psi_i,\phi_j)_{n+m+{1\over2}}+ ... \cr
 &= {\lb G_{{1\over2}} ,{\lb \psi_{i,m},\psi_{j,n}\rb}_{\pm} \rb}_{\pm}\cr
 &= (m-d_i+{\scriptstyle{1\over2}}){\lb
\phi_{i,m+{1\over2}},\psi_{j,n}\rb}_{\pm} +
    (n-d_j+{\scriptstyle{1\over2}}){\lb
\psi_{i,n+{1\over2}},\phi_{j,m}\rb}_{\pm},
} $$
where the dots stand for fields which are linearly
independent from ${\cal N}(\phi_i,\psi_j)$ and ${\cal N}(\psi_i,\phi_j)$.
Because $\coc^{{\cal N}(\phi_i,\psi_j)}_{G\Nuiuj}$ and
$\coc^{{\cal N}(\psi_i,\phi_j)}_{G\Nuiuj}$ are nonzero
and in the commutators in the last line the field
${\cal N}(\phi_i,\psi_i)$ does not occur
$\cocvivjNuiuj$ has to vanish.
\mn
{\bf Remark 3.3:} We use the following normalization
in our calculations instead of (3.4):
$$ d_{\phi_i\phi_i} = { c\over{d(\phi_i)}} \ \ \
{\rm{and}} \ \ \ d_{\psi_i\psi_i} = {c\over{d(\psi_i)}} \eqno({\rm 3.4'})$$
While the normalization $(3.4)$ is more convenient for proofs this one
is more convenient for practical calculations (in this normalization the
$d$-matrices almost factorize completly into linear factors
(cf. appendix A and C) ).
\pano
Now equations $(3.5)$ and $(3.6)$ have to be modified slightly.
One obtains e.g.:
$$C^{\psi_i}_{G\phi_i} = {{2d_i+1}\over{2d_i}} (-1)^{2d_i+1}
C^{\phi_i}_{G\psi_i}
   \ \ \ \ \ {\rm{ and }} \ \ \ \ \ \  {\bigl(
C^{\psi_i}_{G\phi_i}\bigr)}^2 = (-1)^{2d_i+1} (2d_i+1) \eqno({\rm 3.5'})$$
$$  {\bigl( C^{\phi_k}_{{\psi_i}{\phi_j}}\bigr) }^2 = {(2d_k)}^2
{{(-1)^{2d_i+2d_k}(2d_k+1)}\over{2d_i+1}}{\bigl( \Cijk\bigr) }^2 \eqno({\rm
3.6'}) $$
\mn
After this general setup we present an algorithm
for the explicit constuction of $\sw$-algebras.
Let $\cal A$ = $\{G,L,\phi_1,\psi_1,...,\phi_n,\psi_n\}$ be a
simple set and $\Phi_i=\phi_i+\theta\psi_i$ super primaries of
dimension $d_i$.
\pano
Then one proceeds as follows:
\pano
First write down all linearly independent NOP's which may appear in the
commutator of the simple additional fields. The algorithm used is based
on the following facts. Let $\phi$ be a quasiprimary NOP of dimension
$\delta$. Then the quasiprimary projection of
$\partial\phi = \lb L_1,\phi \rb$
is zero and leads to a linear equation in quasiprimary NOPs
of dimension $\delta+1$. Such equations can be used to reduce the set
of all possible NOPs of dimension $\delta+1$ to a basis.
Formally, this can be described by assigning to each quasiprimary field
of dimension $\delta$ a coloured partition $p(\delta)$. The linear
equation
obtained by the quasiprimary projection can be written in terms
of partitions of $\delta+1$. Thus the main step of the algorithm is to
delete for any $p(\delta)$ an arbitrary $p(\delta+1)$ that can be obtained
by adding 1 to some element of $p(\delta)$.
For details of the algorithm as well as for a proof and examples we
refer the reader to $\q{\flo}$.
\mn
Secondly, calculate all structure constants appearing in these commutators.
The structure constants connecting three additional super primary
fields remain as free parameters.
\mn
Finally, one has to check all Jacobi identities with additional
super primary fields only.
This will in general lead to conditions for the free coupling constants
and the central charge.
\mn\mn
Note that the coupling constants containing one $G$ are determined by
(3.5) and the coupling constants between the superconformal families
are determined by  a single coupling constant (cf. (3.6)).
\pano
In order to check the validity of the Jacobi identities
it is sufficient to ensure that all factors in front of the primary fields
vanish.
Primary fields include the additional simple fields and perhaps
fields built from NOPs of two additional fields and $G,L$.
If the coupling constants of the additional simple fields vanish,
such NOPs cannot turn up in the Jacobi identities and
one has to check the factors in front of the additional simple
fields only.
If some coupling constants are nonzero one has to calculate which
primary projections of NOPs of additional simple fields do not vanish
and to ensure that the coefficients in front of these fields are zero.
This is primarily a practical problem since the  calculation of these
factors is very complicated.
Therefore we  checked the factors in front of the simple fields only.
In fact, there is no example showing
that these conditions are not sufficient  (in all previously known
cases we obtained the same results). Indeed there are some examples
where non-simple primary fields occur in the Jacobi identities and
do not lead to further restrictions
(see for example $\sw({3\over2},2)$ in the next chapter).
\mn
For the special case of $\sw$-algebras with two and three generators
this general outline reduces to the following.
\pano
While for
$\sw$-algebras with two generators there remain two free
parameters ($c$ and the self-coupling) there are
five free parameters ($c$ and four coupling constants)
for $\sw$-algebras with three generators.
The coupling constants $C^{j}_{ii}$ vanish if
$d_j \in 2\BN+1$ or $d_j \in 2\BN+{1\over2}$ (cf. remark 3.2).
In almost all calculations we verified that the Jacobi identities
with $L$ or $G$ involved are satisfied automatically,
if one uses (3.5) and (3.6).
In one superconformal family  four Jacobi identities  remain
to be checked:
$$ \eqalign{
  {\lb {\lb \phi_m , \phi_n \rb}_{\pm} ,\phi_k \rb }_{\pm,cycl.} = 0 \cr
  {\lb {\lb \phi_m , \phi_n \rb}_{\pm} ,\psi_k \rb }_{\pm,cycl.} = 0 \cr
  {\lb {\lb \psi_m , \psi_n \rb}_{\pm} ,\phi_k \rb }_{\pm,cycl.} = 0 \cr
  {\lb {\lb \psi_m , \psi_n \rb}_{\pm} ,\psi_k \rb }_{\pm,cycl.} = 0
}$$
For $\sw({3\over2},d_1,d_2)$-algebras there are six Jacobi identities
involving two fields of one additional superconformal family and one
of the second family, such that alltogether 20 Jacobi identities
have to be verified.
\pano
We denote the coupling of $\Phi_i$ with $\Phi_j$ to $\Phi_k$
by $C^{\Phi_k}_{\Phi_i\Phi_j}$, where
$$ C^{\Phi_k}_{\Phi_i\Phi_j} =
   \cases{ C^{\phi_k}_{\phi_i\phi_j} \  , \ d(\Phi_k) \in \BN \cr
           C^{\psi_k}_{\phi_i\phi_j} \  , \ d(\Phi_k) \in \BN +{1\over2}
         } \eqno({\rm 3.6}) $$
\mn
If there exist null fields for a possible discrete value of the
central charge one has to ensure that either all further calculations
are performed without these null fields and fixed central charge or
that they are performed generically.
\mn
\leftline{{\bf 4. Explicit results on $\bf \SW({3\over2},$d)-algebras}}
\mn
In this chapter we present our results on explicit constructions of
several $\sw$(${3\over2},d$)-algebras.
All symbolic calculations have been performed in
$\rm{MATHEMATICA^{TM}}$ and REDUCE while
for the time consuming  commutation operations and the expansion of
NOP's a special C-program had to be used.
\pano
Since the values of the coupling constants do not depend on the special
choice of the basis in the space of quasiprimary fields we will
in general omit the coupling constants involving non-simple fields.
A special choice of a quasiprimary basis is contained in
appendix A and B.
In appendix C we list the Kac-determinants of states  involving
exactly one additional field.
\mn
The cases ${3\over2} \le d \le {7\over2}$ were treated
by  J.Figueroa-O'Farrill et al. $\q{\jos}$.
\mn
\leftline{{$\bf \SW({3\over2},{3\over2}) $ }}
This algebra is a super Lie algebra with the followig commutation
relations:
$$\eqalign{ {\lb \phi_{m},\phi_{n} \rb}_{+} &=
C^{\psi}_{\phi\phi}\psi_{m+n} +2L_{m+n} +
{c\over3}(m^{2}-{1\over4})\delta_{m+n,0}   \cr
            {\lb \phi_{m},\psi_{n} \rb}\  &= {1\over3}(n-2m)\bigl(
C^{\phi}_{\phi\psi}\phi_{m+n} + {3\over2} G_{m+n}\bigr) \cr
            {\lb \psi_{m},\psi_{n} \rb}\  &=
{n-m\over2}C^{\psi}_{\psi\psi}\psi_{m+n} +(n-m)L_{m+n} +
{c\over12}(n^{3}-n)\delta_{m+n,0}
           }$$
where $ C^{\psi}_{\phi\phi} = C^{\psi}_{\psi\psi} \ \ {\rm and} \ \
C^{\phi}_{\phi\psi} = {3\over4}C^{\psi}_{\psi\psi} $
\mn
Thus $\sw$(${3\over2},{3\over2}$) exists generically
and the coupling constant
$ C^{\psi}_{\psi\psi} $ is
a free parameter. Note that one may choose a new basis, such that the
resulting commutators define the direct sum of two
super Virasoro algebras.
This case was discussed in $\q{\jos}$ and is similar to the conformal one
treated
by Zamolodchikov $\q{\zam}$.
\mn
The cases $d = 2,{5\over2}$ with vanishing self-coupling have been
investigated in detail by T. Inami et al. $\q{\ina}$.
\mn
\leftline{{$ \bf \SW({3\over2},2)$ }}
This is the first case where a $\sw$-algebra does not close linearly.
The commutation relations are given by:
$$\eqalign{
  {\lb \phi_{m},\phi_{n} \rb}\ = \
&{n-m\over2}C^{\phi}_{\phi\phi}\phi_{m+n} +(n-m)L_{m+n} +
{c\over12}(n^{3}-n)\delta_{m+n,0}   \cr
  {\lb \phi_{m},\psi_{n} \rb}\ = \ &p_{2,{5\over2},{7\over2}}(m,n)\bigl( \
C^{\n(\phi,G)}_{\phi\psi}\n(\phi,G)_{m+n} +
C^{\n(L,G)}_{\phi\psi}\n(L,G)_{m+n} \ \bigr) + \cr
&p_{2,{5\over2},{5\over2}}(m,n)C^{\psi}_{\phi\psi}\psi_{m+n} + \cr
&p_{2,{5\over2},{3\over2}}(m,n){\textstyle {3\over5} }
C^{\psi}_{G\phi}G_{m+n}   \cr
  {\lb \psi_{m},\psi_{n} \rb}_{+}\  = \
&p_{{5\over2},{5\over2},4}(m,n)\bigl( \
C^{\n(\phi,\phi)}_{\psi\psi}\n(\phi,\phi)_{m+n} +
    C^{\n(\psi,G)}_{\psi\psi}\n(\psi,G)_{m+n} + \cr
                                        & \ \ \ \ \ \ \ \ \ \ \ \ \ \ \ \ \
C^{\n(\phi,L)}_{\psi\psi}\n(\phi,L)_{m+n} +
C^{\n(G,\de{G})}_{\psi\psi}\n(G,\de{G})_{m+n} + \cr
                                        & \ \ \ \ \ \ \ \ \ \ \ \ \ \ \ \ \
C^{\n(L,L)}_{\psi\psi}\n(L,L)_{m+n} \ \bigr) + \cr
&p_{{5\over2},{5\over2},2}(m,n)\bigl( \  C^{\phi}_{\psi\psi}\phi_{m+n} +
     2 L_{m+n} \ \bigr) + \cr
                                        &{2c\over5} { n+{3\over2} \choose
4}\delta_{m+n,0}
          }$$
where $ \ C^{\phi}_{\psi\psi} ={2\over5} C^{\phi}_{\phi\phi} \ \ , \ \
C^{\psi}_{\phi\psi} = {1\over2}C^{\phi}_{\phi\phi}$\
{\rm{and }}
$$ {\bigl( C^{\phi}_{\phi\phi}\bigr) }^{2} = { 4{(5c+6)}^{2} \over
(4c+21)(15-c) }. $$
We have verified explicitly that the coupling constant
$C^{\n(\phi,\phi)}_{\psi\psi}$ is zero.
In general, the field $\n(\phi,\phi)$ cannot be neglected since it
contributes to the $d$-matrix.
Only for $ c = -{6\over5}$ the self-coupling vanishes and all fields
containing $\phi$ or $\psi$ can be ignored. In this case the fields
${\cal N}(\phi,\phi)$ and ${\cal N}(\psi,\phi)$ have vanishing primary
projection. Generically, this algebra contains a super primary field of
dimension 4 which is quadratic in the simple super primary field.
We have verified that if this field occurs in a Jacobi identity the
factor in front of it vanishes (cf. remark at the end of chapter 3).
\pano
Thus the algebra exists generically, which is not surprising since
the classical counterpart is the symmetry algebra of the super Toda
theory corresponding to the super Lie algebra $Osp(3 \vert 2)$.
\mn
\leftline{{$\bf \SW({3\over2},{5\over2}) $ }}
The self-coupling vanishes and only two $c$ values are possible:
$$ c = -{5\over2} , {10\over7} $$
\mn
\leftline{{$\bf\SW({3\over2},3)$  }}
Here the self-coupling is zero again and the allowed $c$ values are:
$$ c = -{45\over2},-{27\over7},{5\over4} $$
\mn
\leftline{{$\bf\SW({3\over2},{7\over2}) $ }}
This is a case with non-vanishing self-coupling and consistency is obtained
for:
\mn
\centerline{
\vbox{ \offinterlineskip
\def\tablespace{ height2pt&\omit&&\omit&\cr }
\def\tablerule{ \tablespace
                \noalign{\hrule}
                \tablespace      }
\hrule
\halign{&\vrule#&
  \strut\quad\hfil#\hfil\quad\cr
\tablespace
& $c$\hfil && ${\bigl(\cowww\bigr) }^2$ &\cr
\tablerule
\tablerule
& ${7\over5}$       && ${146016\over79135} $               &\cr \tablerule
& $-{17\over11}$    && $-{45009216\over523393} $           &\cr \tablespace
}
\hrule}
}
\mn
In the sequel we present new $\sw$(${3\over2},d$)-algebras
with $d$ up to 7.
\mn
\leftline{{$\bf\SW({3\over2},4) $ }}
While for $c = -{20\over3}$ the algebra $\sw$(${3\over2},4$) is consistent
for vanishing self-coupling, it is also consistent with nonzero
self-coupling for four values of $c$:
\mn
\centerline{
\vbox{ \offinterlineskip
\def\tablespace{ height2pt&\omit&&\omit&\cr }
\def\tablerule{ \tablespace
                \noalign{\hrule}
                \tablespace      }
\hrule
\halign{&\vrule#&
  \strut\quad\hfil#\hfil\quad\cr
\tablespace
& $c$\hfil && ${\bigl(\cowww\bigr)}^2$ &\cr
\tablerule
\tablerule
& $-{185\over4}$   && ${355359375\over161644468}$      &\cr \tablerule
& $-13$            && ${427063\over4247}$              &\cr \tablerule
& $-{21\over2}$    && $-{508369\over2499}$             &\cr \tablerule
& $-{120\over13}$  && $-{273043750000\over7804122111}$ &\cr \tablerule
& $-{20\over3}$    && 0                                &\cr \tablespace}
\hrule}
}
\mn
Only in the case $c = -{120\over13}$ the field $\n$($\phi,\phi)$
has a nontrivial primary projection and even in this case the
coupling constant  $\CvvNuu$ vanishes.
This means that there is a
primary field of dimension 8 contained in the algebra at
$c = -{120\over13}$.
In this case the mentioned field contributes to the $d$ matrix.
\mn
\leftline{{$\bf\SW({3\over2},{9\over2})$  }}
This algebra exists only with vanishing self-coupling for three values of
the central charge:
$$ c = -{69\over2},-{81\over10},{4\over11}  $$
\mn
\leftline{{$\bf\SW({3\over2},5)$  }}
Analogously to $\sw({3\over2},3)$ the self-coupling has to vanish.
Consistency implies here $ c = -{105\over11} $.
\mn
\leftline{{$\bf\SW({3\over2},{11\over2})$ }}
The algebra $\sw$(${3\over2},{11\over2}$) is consistent for
five values of $c$:
\mn
\centerline{
\vbox{ \offinterlineskip
\def\tablespace{ height2pt&\omit&&\omit&\cr }
\def\tablerule{ \tablespace
                \noalign{\hrule}
                \tablespace      }
\hrule
\halign{&\vrule#&
  \strut\quad\hfil#\hfil\quad\cr
\tablespace
& $c$\hfil && ${\bigl(\cowww\bigr)}^2$ &\cr
\tablerule
\tablerule
& ${10\over7}$        && ${844918800\over34554863}$           &\cr
\tablerule
& ${11\over40}$       && $-{91376712973\over3514069779}$      &\cr
\tablerule
& $-{155\over19}$     && ${394077798400\over181920009}$       &\cr
\tablerule
& $-{705\over8}$      && $-{4322235328125\over3075066706603}$ &\cr
\tablerule
& $-{5\over13}$       && 0                                    &\cr
\tablespace}
\hrule}
}
\mn
\leftline{{$\bf\SW({3\over2},6)$  }}
This algebra is consistent with non-vanishing self-coupling for
\mn
\centerline{
\vbox{ \offinterlineskip
\def\tablespace{ height2pt&\omit&&\omit&\cr }
\def\tablerule{ \tablespace
                \noalign{\hrule}
                \tablespace      }
\hrule
\halign{&\vrule#&
  \strut\quad\hfil#\hfil\quad\cr
\tablespace
& $c$\hfil && ${\bigl(\cowww\bigr)}^2$ &\cr
\tablerule
\tablerule
& $-{33\over2}$           && ${6309688448\over3137409}$
&\cr \tablerule
& $-18$                   && $-{16916841216\over17064005}$
&\cr \tablerule
& $-{2241\over20}$        && ${788850108780331059\over143584629321555025}$
&\cr \tablespace }
\hrule}
}
\mn
With vanishing self-coupling it exists for:
$$ c = -{93\over2},-{162\over13},{27\over20} $$
The field $\n(\phi,\phi)$ has a non-vanishing primary projection only
for $c = -18$ and $-{2241\over20}$. In these cases $\CvvNuu$ is zero.
\mn
\leftline{{$\bf\SW({3\over2},{13\over2})$  }}
This algebra exists only for zero self-coupling and $ c = -{195\over14} $.
\mn
\leftline{{$\bf\SW({3\over2},7)$  }}
Finally we discuss the algebra obtained by adding a super primary
field of dimension 7. Since 7 is odd the self-coupling vanishes.
For this algebra we have checked
the condition $(\phi_m,\phi_n,\phi_k,\phi_{m+n+k}) = 0$ only.
The only allowed $c$ values are:
$$ c = -{77\over5},-{13\over8} $$
In the next section we will give some arguments that for these values
of the central charge the algebra should indeed exist.
\eject
\leftline{{\bf 5. Structure of $\bf\SW({3\over2},d)$-algebras }}
\mn
In this chapter we summarize our results concerning
$\sw$-algebras with two generators.
We will show that in perfect analogy to $\w$($2,\delta)$-algebras
for most values of the central charge
a classification into special series is possible.
\mn
$\w (2,\delta)$-algebras can be classified by the
values of the central charge. Until now five different classes
appeared:
\item{$\bullet$}
  the generically existing algebras related to simple Lie algebras
  (see e.g. $\q{\bo1}$),
\item{$\bullet$}
  the algebras related to the ADE classification of
  A. Cappelli et al. $\q{\cap}\q{\blm}$,
\item{$\bullet$}
  the algebras of the (1,s) series $\q{\ka1}$,
\item{$\bullet$}
  the parabolic $\w(2,\delta)$-algebras related to degenerate
  representations of the Virasoro \ \ algebra $\q{\mfl}$ and
\item{$\bullet$}
  some exeptional cases $\q{\blm}\q{\eho}$.
\mn
For $\sw(2,d)$-algebras these five types also exist as will be
shown below.
\mn
The algebras $\sw$(${3\over2},d$) with ${3\over2} \le d \le 7$
exist for the following values of the central charge:
\mn
\centerline{
\vbox{ \offinterlineskip
\def\tablespace{ height2pt&\omit&&\omit&\cr }
\def\tablerule{ \tablespace
                \noalign{\hrule}
                \tablespace      }
\hrule
\halign{&\vrule#&
  \strut\quad\hfil#\hfil\quad\cr
\tablespace
& $d$\hfil && $c$ &\cr
\tablerule
\tablerule
& ${3\over2}$  && generic                            &\cr \tablerule
& 2            && generic                            &\cr \tablerule
& ${5\over2}$  && $-{5\over2}$,${10\over7}$          &\cr \tablerule
& 3            && $-{45\over2}$,$-{27\over7}$,
                  ${5\over4}$                      &\cr \tablerule
& ${7\over2}$  && $-{17\over11}$,${7\over5}$         &\cr \tablerule
& 4            && $-{185\over4}$,$-13$,$-{21\over2}$,
                  $-{120\over13}$,$-{20\over3}$      &\cr \tablerule
& ${9\over2}$  && $-{69\over2}$,$-{81\over10}$,
                  ${4\over11}$                       &\cr \tablerule
& 5            && $-{105\over11}$                    &\cr \tablerule
& ${11\over2}$ && $-{705\over8}$,$-{155\over19}$,
                  $-{5\over13}$,${11\over40}$,
                  ${10\over7}$                       &\cr \tablerule
& 6            && $-{2241\over20}$,$-{93\over2}$,
                  $-18$,$-{33\over2}$,$-{162\over13}$,
                  ${27\over20}$                      &\cr \tablerule
& ${13\over2}$ && $-{195\over14}$                    &\cr \tablerule
& 7            && $-{77\over5}$,$-{13\over8}$        &\cr \tablespace}
\hrule}
}
\mn
First, the generically existing algebras $\sw({3\over2},{3\over2})$
and $\sw({3\over2},2)$ are closely related to the Lie super algebras
$Osp(2 \vert 1)$ and $Osp(3 \vert 2)$, respectively $\q{\ina}$.
\mn\mn
Secondly, all super minimal $c$ values can be related to the ADE
classification of modular invariant partition functions of A. Cappelli
et al. $\q{\cap}$. To be more specific, we need
the minimal models of the super Virasoro
algebra. Therefore we list the values of
the central charge and the corresponding superconformal dimensions
$\q{\ber}\q{\eic}\q{\mus}\q{\god}$.
$$ \eqalign{
 c(p,q) &= {3\over2}( 1 - { {2(p-q)}^2\over pq } ) \ \ {\rm with \ either}
\ \
   p,q \ \in\  \BN \  , \ p,q \ \ \ {\rm coprime}
   \ \ {\rm and} \ \ \ p+q \in 2\BN \cr
 & \ \ \ \ \ \ \ \ \ \ \ \ \ \ \ \ \ \ \ \ \ \ \ \ \ \ \ \
{\rm or} \ \ \ \ \ \ \ \ \ \ \ \ \ \
   p,q \in 2\BN \ , \ {p\over2},{q\over2} \ \ {\rm coprime}
   \ \ {\rm and} \ \ {p\over2}+{q\over2} \not\in 2\BN \cr
 h(r,s) &= { {(rp-qs)}^2-{(p-q)}^2 \over 8pq } + {1-{(-1)}^{r+s}\over32}
 \ \ \ 1 \le r \le q-1 \ , \  1 \le s \le p-1
 }  \eqno({\rm 5.1}) $$
$r+s$ even yields representations in the Neveu-Schwarz sector
and $r+s$ odd those in the Ramond sector.
\pano
The divisibility conditions in $(5.1)$ ensure that $p,q \in \BN$ with
$p+q \in 2\BN$ are chosen as small as possible.
\mn
Some simple fusion rule arguments suggest the following
\mn
\leftline{{ {\bf Proposition: } For any $d \in \BN$ with
            $d = {1\over8}(p-2)(q-2),\ \  p,q \in 2\BN \ ,\ \
         {p\over2},{q\over2}$  coprime  and }}
\item{}  ${p+q\over2} \in 2\BN+1$
         the algebra \ $\sw$(${3\over2},d$) exists for
         $ \ c = c(p,q)$ \ with vanishing self-coupling.
\mn
Since in this case the central charge belongs to the super
minimal
series and the dimension of the additional super primary field $\Phi$
can be parametrized by $d = h(1,p-1) = h(q-1,1)$,
the well known fusion rules for super Virasoro minimal models
may be applied $\q{\ber}\q{\eic}\q{\mus}$. In terms of superconformal
families they read
$$ \lb \Phi \rb \times \lb \Phi\rb = \lb 1 \rb $$
such that the additional field of dimension $d$ is a
${\grave{\ }}$simple current${\acute{\ }}$ in the sense of $\q{\sch}$.
In this case general arguments suggest that this
${\grave{\ }}$simple current${\acute{\ }}$
can be added to the algebra of the superconformal family of the identity,
although this has been proven only for the unitary case
$\q{\sch}\q{\dop}\q{\fre}\q{\wil}$.
\pano
Many examples with zero self-coupling may be explained by the proposition.
In fact the two $c$ values for which  $\sw({3\over2},7)$ might exist
belong to this series, so one may assume that $\sw({3\over2},7)$
indeed exists for these two values of the central charge.
\mn
By considering the chiral part of the explicit form of a modular
invariant partition function an interpretation
for all super minimal values of the central charge
is possible $\q{\cap}$:
\mn
\centerline{
\vbox{ \offinterlineskip
\def\tablespace{ height2pt&\omit&&\omit&&\omit&\cr }
\def\tablerule{ \tablespace
                \noalign{\hrule}
                \tablespace      }
\hrule
\halign{&\vrule#&
  \strut\quad\hfil#\hfil\quad\cr
\tablespace
& series && $c(p,q) = (p,q)$\hfil && $h(r,s) = (r,s)$ &\cr
\tablerule
\tablerule
& $(D_{2\rho+2},A_{q-1})$ && $(4\rho+2,q)$    && $(q-1,1) =
{1\over2}\rho(q-2)$ &\cr
\tablerule
& $(E_6,A_{\rho-2})$      && $(\rho-1,12)$    && $(7,1) = {\rho-4\over2}$
&\cr
\tablerule
& \omit                   && \omit            && $(5,1) = {\rho-2\over2}$
&\cr
\tablespace
& $(E_{6},D_{\rho+1})$    && $(2\rho,12)$     && $(7,1) = {2\rho-3\over2}$
&\cr
\tablespace
& \omit                   && \omit            && $(11,1) =
{5(\rho-1)\over2}$&\cr
\tablerule
& \omit                   && \omit            && $(11,1) = {\rho-6\over2}$
  &\cr
\tablespace
& $(E_{8},A_{\rho-2})$    && $(\rho-1,30)$    && $(19,1) =
{3(\rho-4)\over2}$ &\cr
\tablespace
& \omit                   && \omit            && $(29,1) =
{7(\rho-3)\over2}$ &\cr
\tablespace}
\hrule}
}
\mn
Note that all fields corresponding to the dimensions
$h(r,s)$ given above are local.
\pano
Because the partition functions related to the first two rows of the table
consist of binomials only, the number of simple fields in
corresponding RCFTs is at most two.
This case is closely related to $\sw$-algebras
with two generators. Indeed the $(D_{2\rho+2},A_{q-1})$ series leads
exactly to the cases explained by our proposition.
Furthermore, note that in the $(E_6,D_{\rho+1})$ and the
$(E_8,A_{\rho-2})$ series the number of simple fields is at most 4.
So there seems to be a connection to $\sw$-algebras with four generators.
Taking into account that the commutators of the super primary fields
corresponding to the first $h$ values cannot contain the other fields
leads to series of $\sw({3\over2},d)$-algebras, too.
For consistency the other super primary fields contained
in these theories cannot be simple.
\mn
There are two examples that fit in the $(E_6,A_{\rho-2})$ series
both with $\cowww \ne 0$. Examples are
$\sw({3\over2},{7\over2})$ at $c = {7\over5}$ and
$\sw({3\over2},{11\over2})$ at ${10\over7}$. This series is realized
only for odd $\rho$ leading to $(p,q)=(\rho-1,12)$ that fullfil the
divisibility condition in $(5.1)$.
These two Algebras can also be related to the
$(E_6,D_{\rho+1})$ series where the dimension $d$ is given by
the second $h$ value in the table.
The $(E_6,D_{\rho+1})$ series is realized for
$\rho = 7,11,13$ as
$\sw({3\over2},{5\over2}),\ \sw({3\over2},{9\over2})$
and $\sw({3\over2},{11\over2})$ with vanishing self-coupling.
Note that only the cases where $\rho$ and 6 are coprime appear.
Since $\sw({3\over2},{5\over2})$ and $\sw({3\over2},{11\over2})$
exist for $c={10\over7}$ and both can be related to $(E_6,D_8)$ there
should be a connection. We have explicitly verified that
$\sw({3\over2},{5\over2})$ contains a super primary
non-simple field of dimension ${11\over2}$ at $c= {10\over7}$.
There is strong evidence  that
this yields a realization of $\sw({3\over2},{11\over2})$ at the
$c$ value mentioned because we have calculated the self-coupling
of one component of the non-simple super primary field.
Its value is exactly that obtained
by the direct construction of $\sw({3\over2},{11\over2})$.
For details we refer to appendix D.
A similar construction where $\sw({3\over2},{7\over2})$ is realized
as a subalgebra of $\sw({3\over2},{3\over2})$ at $c = {7\over5}$
has been carried out by K. Hornfeck $\q{\hor}$.
\mn
The $(E_8,A_{\rho-2})$ series is realized only for $\rho = 17$ as
$\sw({3\over2},{11\over2})$ at $c={11\over40}$.
\mn
For $12 \le \rho  \le 16$
the resulting pair $(\rho-1,30)$ does not  belong to the $(E_8,A_{\rho-1})$
series since  the divisibility condition in $(5.1)$ is not satisfied.
The next possible value for $\rho$ is $23$ and the corresponding
dimension ${17\over2}$.
\mn
\mn
Thirdly, note that $\sw({3\over2},d)$ exists for $c = c(1,s)$, $s$ odd
and $d = {2s-1\over2}$.
This case is similar to the conformal one treated by H.G. Kausch for
$\w$-algebras
with two generators.
In $\q{\ka1}$ a free field construction was presented and one can
hope that similar techniques lead to a free field realization in the
supersymmetric case. We will call this series the $(1,s)$-series.
\mn  \mn
Fourthly, all $\sw({3\over2},d)$ with $d = {3\over2}n$
or  $d = 2n \ , \ n \in \BN$ exist for
$c = {3\over2}(1-{16\over3}d) = {3\over2}(1-8n) = c^{3}_{d}(n)$
or $c= {3\over2}(1-2d) = {3\over2}(1-4n)=c^{8}_{d}(n)$ respectively.
These cases are related to degenerate representations of the super
Virasoro algebra and will be called parabolic.
The central charge and the dimensions of the primary
fields for the super Virasoro algebra are given by $\q{\eic}$:
$$ \eqalign{
   c &= {3\over2}(1-16{\alpha_0}^2) \cr
   \alpha_{\pm} &= \sqrt{ {\alpha_0}^2+{1\over2} } \pm \alpha_0   \cr
   h_{r,s} &= { {(r\alpha_{+} - s\alpha_{-})}^2 \over4} - {\alpha_{0}}^2
  } \eqno( \rm{5.2} )  $$
Note that  $h_{r,r}  = {\alpha_{0}}^2(r^2-1)$ and
           $h_{r,-r} = {\alpha_{0}}^2(r^2-1)+{1\over2}r^2$.
The series described above can be obtained by adding an
additional field of superconformal dimension $d = h_{2,2}$
or $d = h_{3,3}$ to the superconformal family of the identity.
Note that these fields are local because they obey the locality
condition
$$\epsilon_{{d}{d}} =  e^{2\pi i  {(1-r)}^2 \alpha_0^2} =
                            e^{2\pi i {{(1-r)}^2\over r^2-1} d } =
                            \pm 1$$
if one assumes $2{ {(r-1)}^2\over r^2-1} d \in \BN$. Taking into account
that the fields corresponding to $h_{2,2}$ or $h_{3,3}$ are either
bosonic or fermionic leads to the possible values of $d$ given above.
For $r \ge 4$ one would
not get a $\sw$-algebra with only two generators.
These algebras lead to rational theories with effective central charge
${\tilde c} = {3\over2}$ $\q{\eh1}$.
For the similar case of $\w(2,\delta)$-algebras the characters and
$S$-matrices have
been calculated and the generalization to the super case is
straightforward $\q{\mfl}$.
\mn \mn
Finally, some algebras remain which do not belong to any of the series
described above. All of them have non-vanishing self-coupling.
The investigation of the representation theory of these
algebras leads to a better understanding. One obtains e.g.
that the effective central charge is greater than ${3\over2}$
$\q{\eh1}$.
\mn
Finally we list all $\sw({3\over2},d)$-algebras with
${3\over2} \le d \le 7$ and show how they fit into these patterns.
\vfill
\eject
\centerline{
\vbox{ \offinterlineskip
\def\tablespace{   height2pt&\omit&&\omit&&\omit&&\omit&&\omit&\cr }
\def\tablespaceno{ height2pt&\omit&&\multispan3 &&\omit&&\omit&\cr }
\def\tablerule{ \tablespace
                \noalign{\hrule}
                \tablespace      }
\hrule
\halign{&\vrule#&
  \strut\quad\hfil#\hfil\quad\cr
\tablespaceno
& $d$     && \multispan3 \hfil $c = c(p,q)$ \hfil    && $d = h(r,s)$  &&
series            &\cr
\tablespaceno
\noalign{\hrule}
\tablespace
& \omit   && $\cowww = 0$  && $\cowww \ne 0$         && \omit         &&
\omit             &\cr
\tablerule
\tablerule
& ${3\over2}$  && generic                 && generic     && \omit   &&
\omit                  &\cr
\tablerule
& 2            && $(10,4) = -{6\over5}  $ &&\omit        && (3,1)   &&
$(D_6,A_3)$            &\cr
\tablespace
& \omit        &&\omit && generic                        && \omit   &&
\omit                  &\cr
\tablerule
& ${5\over2}$  && $(3,1) = -{5\over2}   $ &&\omit        && (3,1)   &&
$(1,s)$                &\cr
\tablespace
& \omit        && $(14,12) = {10\over7} $ &&\omit        && (5,1)   &&
$(E_6,D_8)$            &\cr
\tablerule
& \omit        && $(14,4) = -{27\over7} $ &&\omit        && (3,1)   &&
$(D_8,A_3)$            &\cr
\tablespace
& 3            && $(6,8) = {5\over4}    $ &&\omit        && (7,1)   &&
$(D_4,A_7)$            &\cr
\tablespace
& \omit        && $c_{d}^3(2) =-{45\over2}$ &&\omit      && (2,2)   &&
parabolic              &\cr
\tablerule
& ${7\over2}$  && \omit && $(10,12) = {7\over5}$         && (7,1)   &&
$(E_6,A_{9})$
  &\cr \tablespace
& \omit        && \omit && $-{17\over11}$                && \omit   && ?
                  &\cr
\tablerule
& \omit        && $(18,4) = -{20\over3} $ &&\omit        && (3,1)   &&
$(D_{10},A_3)$         &\cr
\tablespace
& \omit        && \omit &&$c_{d}^8(2)=-{21\over2}$       && (3,3)   &&
parabolic              &\cr
\tablespace
& 4            && \omit && $-{185\over4}$                && \omit   && ?
                  &\cr
\tablespace
& \omit        && \omit && $-13$                         && \omit   && ?
                  &\cr
\tablespace
& \omit        && \omit && $-{120\over13}$               && \omit   && ?
                  &\cr
\tablerule
& \omit        && $(5,1) = -{81\over10} $ &&\omit        && (3,1)   &&
$(1,s)$                &\cr
\tablespace
& ${9\over2}$  && $(22,12) = {4\over11} $ &&\omit        && (5,1)   &&
$(E_6,D_{12})$         &\cr
\tablespace
& \omit        && $c_{d}^3(3)=-{69\over2}$&&\omit        && (2,2)   &&
parabolic              &\cr
\tablerule
& 5            && $(22,4) = -{105\over11}$&&\omit        && (3,1)   &&
$(D_{12},A_3)$         &\cr
\tablerule
& \omit        && $(26,12) = -{5\over13} $&&\omit        && (5,1)   &&
$(E_6,D_{14})$         &\cr
\tablespace
& \omit        &&\omit && $(14,12) = {10\over7}$         && (7,1)   &&
$(E_6,A_{13})$         &\cr \tablespace
& ${11\over2}$ &&\omit && $(16,30) = {11\over40}$        && (11,1)  &&
$(E_8,A_{15})$         &\cr \tablespace
& \omit        &&\omit && $-{705\over8}$                 && \omit   && ?
                  &\cr
\tablespace
& \omit        &&\omit && $-{155\over19}$                && \omit   && ?
                  &\cr
\tablerule
& \omit        && $(26,4) = -{162\over13}$&&\omit        && (3,1)   &&
$(D_{14},A_3)$         &\cr
\tablespace
& 6            && $(10,8) = {27\over20}  $&&\omit        && (7,1)   &&
$(D_{6},A_7)$          &\cr
\tablespace
& \omit        && $c_{d}^3(4)=-{93\over2}$&&\omit        && (2,2)   &&
parabolic              &\cr
\tablespace
& \omit        &&\omit &&$c_{d}^8(3)=-{33\over2}$        && (3,3)   &&
parabolic              &\cr
\tablespace
& \omit        &&\omit &&$-18$                           && \omit   && ?
                  &\cr
\tablespace
& \omit        &&\omit &&$-{2241\over20}$                && \omit   && ?
                  &\cr
\tablerule
& ${13\over2}$ && $(7,1) = -{195\over14} $&&\omit        && (3,1)   &&
$(1,s)$                &\cr
\tablerule
& 7            && $(30,4) = -{77\over5}  $&&\omit        && (3,1)   &&
$(D_{16},A_3)$         &\cr
\tablespace
& \omit        && $(6,16) = -{13\over8}  $&&\omit        && (15,1)  &&
$(D_{4},A_{15})$       &\cr
\tablespace}
\hrule}
}
\mn
\leftline{{\bf 6. Explicit results about $\bf\SW({3\over2},
{\textstyle d_1,d_2})$-algebras  }}
\mn
In this charpter we present our results about the construction of
$\sw$-algebras with two additional generators.
For these algebras one has to check the validity of 20 Jacobi identities.
For
$\sw({3\over2},d_1,d_2)$ we will denote the two additional super fields
of dimension  $d_i$ by $\Phi^i= \phi^i + \theta\psi^i $.
\mn
\leftline{{$\bf\SW({3\over2},{3\over2},{3\over2}) $ }}
In this case there are five free parameters namely the central charge
and four coupling constants. Consistency implies that the commutators
close linearly in the fields and also the following condition:
$$ {\bigl( C^{\psi^2}_{{\phi^1}{\phi^1}}\bigr)}^2 +
   {\bigl( C^{\psi^1}_{{\phi^2}{\phi^2}}\bigr)}^2 -
   C^{\psi^2}_{{\phi^1}{\phi^1}} C^{\psi^2}_{{\phi^2}{\phi^2}} -
   C^{\psi^1}_{{\phi^2}{\phi^2}} C^{\psi^1}_{{\phi^1}{\phi^1}} - 4 = 0 $$
In complete analogy to the $\sw({3\over2},{3\over2})$-algebra one can
choose a new basis such that the resulting commutators define
the direct sum of three super Virasoro algebras. As described in
$\q{\jos}$ this case can be generalized easily to the extension of
the super Virasoro algebra by $n$ simple fields of dimension ${3\over2}$.
\mn
The following algebra has been studied in $\q{\blu}$
using the covariant approach.
\mn
\leftline{{$\bf\SW({3\over2},{3\over2},2)$  }}
For this algebra five parameters have to be calculated.
The Jacobi identities imply that two sets of solutions exist:
\pano
\leftline{{1.) }}
$$ \eqalign{
   C^{\phi^2}_{\phi^1\phi^1} &= 0  \cr
   C^{\psi^1}_{\phi^1\phi^1} &= { {\bigl(
C^{\psi^1}_{\phi^2\phi^2}\bigr)}^2 - 4 \over
                                  C^{\psi^1}_{\phi^2\phi^2} } \cr
   {\bigl(C^{\phi^2}_{\phi^2\phi^2}\bigr)}^2 &=
  { 4{\bigl(12+10c+3{\bigl(C^{\psi^1}_{\phi^2\phi^2}\bigr)}^2\bigr)}^2
     \bigl(4+{\bigl(C^{\psi^1}_{\phi^2\phi^2}\bigr)}^2\bigr)
    \over
     \bigl(84+16c+21{\bigl(C^{\psi^1}_{\phi^2\phi^2}\bigr)}^2\bigr)
     \bigl(60-4c+15{\bigl(C^{\psi^1}_{\phi^2\phi^2}\bigr)}^2\bigr)
  }
  }$$
\leftline{{2.) }}
$$ \eqalign{
   C^{\psi^1}_{\phi^2\phi^2} &= {1\over2}C^{\psi^1}_{\phi^1\phi^1} \cr
   {\bigl( C^{\psi^1}_{\phi^1\phi^1}\bigr)}^2 &=
   4{\bigl( C^{\phi^2}_{\phi^1\phi^1}\bigr)}^2 -
   4C^{\phi^2}_{\phi^1\phi^1}C^{\phi^2}_{\phi^2\phi^2} - 16  \cr
   C^{\phi^2}_{\phi^1\phi^1} &= {8c\over 10c-27}C^{\phi^2}_{\phi^2\phi^2}
 }$$
This means that in both cases there remain two free parameters.
While the first solution is a trivial one as discussed below the
second solution could be the symmetry algebra of the quantized Toda theory
corresponding to $D(2 \vert 1,\alpha)$.
\mn
\leftline{{$\bf \SW({3\over2},{3\over2},{5\over2})$  }}
Since the coupling constant of the superconformal family of dimension
${3\over2}$ to the superconformal family of dimension
${5\over2}$ and the self-coupling of the latter are zero
there remain three free parameters.
Two types of solutions exist with vanishing
$C^{\psi^2}_{{\phi^1}{\phi^1}}$:
\pano
\leftline{{1.) }}
$$ \eqalign{
   &{\bigl(C^{\psi^1}_{\phi^2\phi^2}\bigl)}^2 = {14c-20\over5} \ \ \ \ \ \
\ \
   C^{\psi^1}_{\phi^1\phi^1}C^{\psi^1}_{\phi^2\phi^2} = {2(7c-20)\over5} \
\ \ c \ne {10\over7} \cr
   {\rm or }\ \ \ \ \ &{\bigl(C^{\psi^1}_{\phi^2\phi^2}\bigr)}^2 =
-{8c+20\over5} \ \ \ \ \ \ \ \ \
                      C^{\psi^1}_{\phi^1\phi^1}C^{\psi^1}_{\phi^2\phi^2} =
-{8(c+5)\over5} \ \ \ c \ne -{5\over2}
}$$
\leftline{{2.) }}
$$ \eqalign{
   {\bigl(C^{\psi^1}_{\phi^2\phi^2}\bigr)}^2 &= -4 \ \ \ \ \ \ \ \ \ \ \ \
   C^{\psi^1}_{\phi^1\phi^1}C^{\psi^1}_{\phi^2\phi^2} = -8
}$$
In both cases the central charge is free.
\mn
For $\sw({3\over2},{3\over2},2)$ and $\sw({3\over2},{3\over2},{5\over2})$
the first solutions can be decomposed into a direct sum.
In general, one has the following structure.
Assume that $\sw({3\over2},d)$ with generators
${\hat {\cal L} }$ and ${\hat \Phi}$
exists for $\hat c$ and self-coupling
${\hat C}^{\hat {\Phi} }_{{\hat\Phi}{\hat\Phi}} = f({\hat c})$.
Then $\sw({3\over2},{3\over2},d)$ exists for generic central charge $c$
and:
$$ \eqalign{
   C^{\psi^2}_{{\phi^1}{\phi^1}} &= 0 \ \ \ \ \ \ \
   {\bigl( C^{\psi^1}_{{\phi^1}{\phi^1}}\bigr) }^2 =
   {4\over{\hat c}}{{(c-2{\hat c})}^2 \over c-{\hat c}} \cr
   {\bigl( C^{\psi^1}_{{\phi^2}{\phi^2}}\bigr) }^2 &= 4{c-{\hat c} \over
{\hat c} } \ \ \ \ \
   {C}^{\Phi}_{\Phi\Phi} =  {c\over{\hat c}}
   {f}\bigl( 4{(4+C^{\psi^1}_{{\phi^2}{\phi^2}})}^{-1} \bigr)
}$$
This is easily seen since a change of basis implies for this solution:
$$ \sw({\textstyle {3\over2},{3\over2}},d)
\cong \sw({\textstyle{3\over2}}) \oplus \sw({\textstyle {3\over2}},d) $$
For the second solution of $\sw({3\over2},{3\over2},{5\over2})$ such a
change of basis becomes singular and is therefore not possible.
It is remarkable that one can choose a linear combination of
the super Virasoro field and the additional super field of dimension
${3\over2}$ such that the resulting field (anti-)commutes with itself.
This field is built up by the sum of the Super Virasoro field and
${ C^{\psi^1}_{\phi^1\phi^1}\over4 }$ times the field $\Phi^1$.
\mn
\leftline{{$\bf\SW({3\over2},2,2)$ }}
Consistency implies fixed central charge
and the following conditions for the four free coupling constants:
$$ c = {3\over2}  $$
$$ C^{\phi^1}_{{\phi^2}{\phi^2}} =  - C^{\phi^1}_{{\phi^1}{\phi^1}} \ \ \
\ \ \
   C^{\phi^2}_{{\phi^1}{\phi^1}} =  - C^{\phi^2}_{{\phi^2}{\phi^2}} $$
$$ {\bigl( C^{\phi^1}_{{\phi^1}{\phi^1}} \bigr)}^2 +
   {\bigl( C^{\phi^2}_{{\phi^2}{\phi^2}} \bigr)}^2 = 2 $$
For this solution the fields ${\cal N}(\phi^1,\phi^1),
{\cal N}(\phi^2,\phi^2)$ and ${\cal N}(\phi^1,\phi^2)$ have no
primary projection and do not contribute to the algebra.
\mn
Obviously, this algebra has an inner $SO(2)$ symmetry realized as a
 rotation
in the space of the super fields of dimension two. Consequently the
solution
is determined by relations that are invariant under the action of $SO(2)$.
Note that a rotation by an angle $\alpha$ in the space of super fields
yields a rotation of angle $3\alpha$ in the space of self-couplings.
\mn
One could speculate that $\sw({3\over2},2,2)$ for $c={3\over2}$ is a
subalgebra of $\sw({3\over2},2,2,{7\over2})$, which should exist
generically and is related to $Osp(4 \vert 4)$ $\q{\del}$.
A very similar case for the $\w(2,4,4)$-algebra which is related to
$SO(8)$ has been treated in $\q{\ka1}$.
\mn
\leftline{{$\bf\SW({3\over2},2,{5\over2})$ }}
For this algebra there exist two consistent sets of solutions:
\pano
\leftline{{1.) }}
$$ \eqalign{
 c &= -15   \cr
 {\bigl( C^{\phi^1}_{{\phi^2}{\phi^2}}\bigr) }^2 &= -{40\over13}   \cr
 {\bigl( C^{\phi^1}_{{\phi^1}{\phi^1}}\bigr) }^2 &= -{1058\over65}  \cr
 C^{\phi^1}_{{\phi^2}{\phi^2}} C^{\phi^1}_{{\phi^1}{\phi^1}} &=
 -{92\over13}
}$$
\leftline{{2.) }}
$$ \eqalign{
 c &= {39\over2}   \cr
 {\bigl( C^{\phi^1}_{{\phi^2}{\phi^2}}\bigr) }^2 &= -88   \cr
 {\bigl( C^{\phi^1}_{{\phi^1}{\phi^1}}\bigr) }^2 &= -{1058\over11}  \cr
 C^{\phi^1}_{{\phi^2}{\phi^2}} C^{\phi^1}_{{\phi^1}{\phi^1}} &= -92
}$$
For $c = -15$ the field ${\cal N}(\phi^1,\phi^2)$ can be written as
a linear combination of the other fields of dimension ${9\over2}$.
This is the only case in which null fields appear and must be
omitted (cf. the remark at the end of chapter 3).
\mn
\leftline{{$\bf\SW({3\over2},{5\over2},{5\over2})$ }}
There is no solution for this algebra.
Analogously to $\w$-algebras a $\sw$-algebra with two additional
super fields and no nonzero coupling constant cannot exist,
since a Jacobi identity of type
${\bigl( {\lb {\lb {\phi^1}_m,{\phi^1}_n \rb}_{\pm},{\phi^2}_k
\rb}_{\pm}\bigr) }_{ cycl.} = 0 $
cannot be satisfied $\q{\blm}$.
\pano
That means $\sw({3\over2},d_1,d_2)$ cannot exist for
$\lb d_1+{1\over2} \rb,\lb d_2+{1\over2} \rb \in 2\BN + 1$.
\mn
\leftline{{$\bf \SW({3\over2},{5\over2},{7\over2})$ }}
In this case two coupling constants are free and one obtains only one set
of solutions with no free parameters:
$$ \eqalign{
   c &= {13\over6}     \cr
   {\bigl(C^{\psi^2}_{\phi^1\phi^1}\bigr)}^2 &= {5208\over979} \cr
   {\bigl(C^{\psi^2}_{\phi^2\phi^2}\bigr)}^2 &= {317709\over60698}\cr
   C^{\psi^2}_{\phi^1\phi^1}C^{\psi^2}_{\phi^2\phi^2} &= -{5166\over979}
}$$
This is an algebra predicted by Schoutens et al. in $\q{\sou}$ by coset
considerations.
It has been shown in $\q{\hor}$ by K. Hornfeck that if this $\sw$-algebra
has a $\w(2,3,4)$ as subalgebra it can at most exist for
$c = {13\over6}$.
Indeed this value is the only possible one.
\mn
\leftline{{\bf 7. Conclusion }}
\mn
Using a non-explicitly covariant approach we have been able to construct
a whole bunch of new $\sw$-algebras with two generators.
Most of the results fit into systematic patterns in complete analogy to
$\w$-algebras. For some new $\sw$-algebras namely
$\sw({3\over2},4)$  at $c = -{185\over4},-{120\over13},-13 $ and
$\sw({3\over2},{11\over2})$  at $c = -{705\over8},-{155\over19} $ and
$\sw({3\over2},6)$  at $c = -18,-{2241\over20} $ an interpretation
has not been found yet. Even the interpretation of
$\sw({3\over2},{7\over2})$ at $c = -{17\over11}$  is still  an
open question  although this algebra has been known for some time.
A study of the possible and physically relevant
highest weight repesentations of these algebras will
lead to a better understanding of these algebras. The methods
explained in $\q{\rva}\q{\eho}$ can be applied to this case with only small
changes and work is in progress $\q{\eh1}$.
\pano
In contrast to $\w(2,\delta)$-algebras no consistent
$\sw({3\over2},d)$-algebras with irrational values of the central
charge appeared. Furthermore, there is only one generically existing
nonlinear $\sw({3\over2},d)$-algebra, namely $\sw({3\over2},2)$.
\pano
For $\sw$-algebras with three generators we have constructed two new
algebras that exist for finitely many $c$ values only.
\mn
Since the difficulties of the transition from $\w$-algebras to
$\sw$-algebras  have not been too hard we are confident that
similar calculations for the super $N=2$ case will be possible
in the near future.
\mn
\leftline{{\bf Acknowledgements  }}
\mn
We would like to thank
W. Nahm, M. Flohr, J. Kellendonk, S. Mallwitz, A. Recknagel,
M. R{\"o}sgen, M. Terhoeven and R. Varnhagen
for many useful discussions. The creative atmosphere
in the institute was an important support  for this work.
\mn
It is a pleasure to thank the Max-Planck-Institut f\"{u}r Mathematik
in Bonn-Beuel especially Th. Berger, S. Mauermann and T. H{\"o}fer
since nearly all calculations have been performed
in their computer centre.
\mn
\eject
\def\detsp{ \hskip 58.0 pt }
\leftline{{\bf Appendix: }}
\bn
In the appendices A, B and C we list all fields and determinants
(of matrices $(d_{ij})$) which occur in the examples considered
in chapter 4. Because fields of different conformal dimensions
are orthogonal, the matrices have block-diagonal form. Furthermore,
all fields in the superconformal family of the identity have
vanishing $(d_{ij})$ with fields of the superconformal family
of the additional super primary field. Thus the determinants factorize
in two terms:
$$ det D_{\delta}^{} = det D_{\delta}^{\lb 1 \rb}
det D_{\delta}^{\lb d \rb} $$
where $\lb d \rb$ stands for the superconformal family of the
additional super primary field  of dimension $d$ and $\delta$ is the
dimension considered. But there is one exception to this rule: if
$d \in 2\BN $ and $\delta = 2d$ the field  $\n(\phi,\phi)$ is involved
and the determinant $det D_{\delta}^{}$ does not factorize.
Because such determinants are very complicated we omit them.
\mn
\normalspace
\leftline{{\bf Appendix A: A basis of quasiprimary fields
               up to dimension $25\over2$ built up by $\cal L$}}
{
\mn
\leftline{$d ={3\over2}$ : 1 field :   $G$}
\leftline{$det D^{\lb1\rb}_{{3\over2}}  =  {2c\over3}$}
\mn
\leftline{$d$ = 2  : 1 field :   $L$}
\leftline{$det D^{\lb1\rb}_{2}    =  {c\over2}$}
\mn
\leftline{$d ={7\over2}$  : 1 field :   $\n(L, G)$}
\leftline{$det D^{\lb1\rb}_{{7\over2}}  =  {1\over12}c(21 + 4c)$}
\mn
\leftline{$d$ = 4  : 2 fields : $\n(G, {\de}G), \ \ \n(L, L)$}
\leftline{$det D^{\lb1\rb}_{4}    =  {1\over60}c^{2}(21 + 4c)(-7 + 10c)$}
\mn
\leftline{$d ={9\over2}$  : 1 field : $\n(L, {\de}G)$}
\leftline{$det D^{\lb1\rb}_{{9\over2}}  = -{2\over35}c(-7 + 10c)$}
\mn
\leftline{$d ={11\over2}$  : 2 fields : $\n(L, \de^{2}G),\ \ \n(\n(L, L),
G)$}
\leftline{$det D^{\lb1\rb}_{{11\over2}} =  {1\over54}c^{2}(11 + c)(21 +
4c)(-7 + 10c)$}
\mn
 \leftline{$d$ = 6  : 4 fields :}
\leftline{$\n(G, {\de}^{3}G),\ \ \n(\n(L, G), {\de}G),\ \ \n(\n(L, L), L),\
\ \n(L, {\de}^{2}L)$}
\leftline{$det D^{\lb1\rb}_{6}    =  {1\over378}c^{4}(11 + c)(21 +
4c)^{2}(-7 + 10c)^{2}(11 + 14c)$}
\mn
\leftline{$d ={13\over2}$  : 2 fields :}
\leftline{$\n(L, {\de}^{3}G),\ \ \n(\n(L, L), {\de}G)$}
\leftline{$det D^{\lb1\rb}_{{13\over2}} =  {16\over847}c^{2}(21 + 4c)(-7 +
10c)(11 + 14c)$}
\mn
\leftline{$d$ = 7  : 1 field :  $\n(\n(L, G), {\de}^{2}G)$}
\leftline{$det D^{\lb1\rb}_{7}    =  -{2\over45}c(21 + 4c)(-7 + 10c)$}
\mn
\leftline{$d ={15\over2}$  : 5 fields :}
\leftline{$\n(\n(G, {\de}G), {\de}^{2}G),\ \  \n(L, {\de}^{4}G),\ \
\n(\n(L, L), {\de}^{2}G),$}
\leftline{$\n(\n(\n(L, L), L), G), \ \ \n(\n(L, {\de}^{2}L), G)$}
\leftline{$det D^{\lb1\rb}_{{15\over2}} =  {64\over27885}(-1 + c)c^{5}(11 +
c)(21 + 4c)^{3}(135 + 8c)(-7 + 10c)^{3}(11 + 14c)$}
\mn
\mn
 \leftline{$d$ = 8  : 7 fields :}
\leftline{$\n(G, {\de}^{5}G) ,\ \ \n(\n(L, G), {\de}^{3}G),\ \ \n(\n(L,
{\de}G), {\de}^{2}G),$}
\leftline{$\n(\n(\n(L, L), G), {\de}G),\ \ \n(\n(\n(L, L), L), L),\ \
\n(\n(L, L), {\de}^{2}L),\ \ \n(L, {\de}^{4}L)$}
\leftline{$det D^{\lb1\rb}_{8}    =  {5184000\over24191167}(-1 + c)c^{7}(11
+ c)^{2}(21 + 4c)^{4}(135 + 8c)(-7 + 10c)^{4}(11 + 14c)^{2}$}
\mn
\mn
\leftline{$d ={17\over2}$  : 4 fields :}
\leftline{$\n(L, {\de}^{5}G),\ \ \n(\n(L, L), {\de}^{3}G),\ \ \n(\n(\n(L,
L), L), {\de}G),\ \ \n(\n(L, {\de}^{2}G), {\de}G)$}
\leftline{$det D^{\lb1\rb}_{{17\over2}} = {36864\over65065}c^{4}(11 + c)(21
+ 4c)^{3}(-7 + 10c)^{3}(11 + 14c)$}
\mn
\mn
 \leftline{$d$ = 9  : 4 fields :}
\leftline{$\n(\n(L, G), {\de}^{4}G),\ \ \n(\n(L, {\de}G), {\de}^{3}G),\ \
\n(\n(\n(L, L), G), {\de}^{2}G),\ \ \n(\n(L, L), {\de}^{3}L)$}
\leftline{$det D^{\lb1\rb}_{9}    = {108\over343}(-1 + c)c^{4}(11 + c)(21 +
4c)^{3}(-7 + 10c)^{3}(11 + 14c)$}
\mn
\mn
\leftline{$d ={19\over2}$  : 9 fields :}
\leftline{$\n(\n(G, {\de}G), {\de}^{4}G) ,\ \ \n(\n(\n(L, G), {\de}G),
{\de}^{2}G),$}
\leftline{$\n(L, {\de}^{6}G),\ \ \n(\n(L, L), {\de}^{4}G),\ \ \n(\n(\n(L,
L), L), {\de}^{2}G),$}
\leftline{$\n(\n(L, {\de}^{2}L), {\de}^{2}G),\ \ \n(\n(\n(\n(L, L), L), L),
G),\ \ \n(\n(\n(L, L),{\de}^{2}L), G),$}
\leftline{$\n(\n(L, {\de}^{4}L), G)$}
\leftline{$det D^{\lb1\rb}_{{19\over2}} = {8599633920\over53202877}(-1 +
c)^{2}c^{9}(11 + c)^{3}(21 + 4c)^{6}  $}
\leftline{$\detsp  (114 + 5c)(135 + 8c)(-7 + 10c)^{6}(11 + 14c)^{3} $}
\mn
\mn
\leftline{$d$ = 10  : 12 fields :}
\leftline{$\n(G, {\de}^{7}G),\ \ \n(\n(L, G), {\de}^{5}G),\ \ \n(\n(L,
{\de}G), {\de}^{4}G),\ \ \n(\n(L, {\de}^{2}G), {\de}^{3}G),$}
\leftline{$\n(\n(\n(L, L), G), {\de}^{3}G),\ \ \n(\n(\n(L, L), {\de}G),
{\de}^{2}G),\ \ \n(\n(\n(\n(L, L), L), G), {\de}G),$}
\leftline{$\n(\n(\n(L, {\de}^{2}L), G), {\de}G),\ \ \n(\n(\n(\n(L, L), L),
L), L),\ \ \n(\n(\n(L, L), L), {\de}^{2}L)$}
\leftline{$\n(\n(L, L), {\de}^{4}L),\ \ \n(L, {\de}^{6}L)$}
\leftline{$det D^{\lb1\rb}_{10}   =
{1011316948992000000\over762819593107}(-1 + c)^{2}c^{12}(11 + c)^{4}(21 +
4c)^{8}(114 + 5c)$}
\leftline{$\detsp (135 + 8c)^{2}(-7 + 10c)^{8}(11 + 14c)^{4}(95 + 22c)$}
\mn
\mn
\leftline{$d ={21\over2}$  : 10 fields :}
\leftline{$\n(\n(G, {\de}G), {\de}^{5}G),\ \ \n(\n(\n(L, G), {\de}G),
{\de}^{3}G),\ \ \n(L, {\de}^{7}G), \n(\n(L, L), {\de}^{5}G),$}
\leftline{$\n(\n(\n(L, L), L), {\de}^{3}G),\ \ \n(\n(L, {\de}^{2}L),
{\de}^{3}G),\ \ \n(\n(\n(\n(L, L), L), L), {\de}G),$}
\leftline{$\n(\n(\n(L, L), {\de}^{2}L), {\de}G),\ \ \n(\n(L, {\de}^{4}L),
{\de}G),\ \ \n(\n(\n(L, L), {\de}^{3}L), G)$}
\leftline{$det D^{\lb1\rb}_{{21\over2}} =
{477199206497805926400000000\over65518365270271857229}(-1 + c)^{2}c^{10}(11
+ c)^{3}(21 + 4c)^{7}$}
\leftline{$\detsp (135 + 8c)(-7 + 10c)^{7}(11 + 14c)^{4}(95 + 22c)$}
\mn
\mn
 \leftline{$d$ = 11  : 9 fields :}
\leftline{$\n(\n(L, G), {\de}^{6}G),\ \ \n(\n(L, {\de}G), {\de}^{5}G),\ \
\n(\n(L, {\de}^{2}G), {\de}^{4}G),$}
\leftline{$\n(\n(\n(L, L), G), {\de}^{4}G),\ \ \n(\n(\n(L, L), {\de}G),
{\de}^{3}G),\ \ \n(\n(\n(\n(L, L), L), G), {\de}^{2}G),$}
\leftline{$\n(\n(\n(L, {\de}^{2}L), G), {\de}^{2}G),\ \ \n(\n(\n(L, L), L),
{\de}^{3}L),\ \ \n(\n(L, L), {\de}^{5}L)$}
\leftline{$det D^{\lb1\rb}_{11}   =  -{28179280429056\over2215547}(-1 +
c)^{2}c^{9}(11 + c)^{3}(21 + 4c)^{7}(135 + 8c)(-7 + 10c)^{7}(11 +
14c)^{4}$}
\mn
\leftline{$d ={23\over2}$ :  16 fields :}
\leftline{$\n(\n(G, {\de}G), {\de}^{6}G),\ \ \n(\n(\n(L, G), {\de}G),
{\de}^{4}G),\ \ \n(\n(\n(L, G), {\de}^{2}G), {\de}^{3}G)$}
\leftline{$\n(L, {\de}^{8}G),\ \ \n(\n(\n(\n(L, L), G), {\de}G),
{\de}^{2}G),\ \ \n(\n(L, L), {\de}^{6}G),$}
\leftline{$\n(\n(\n(L, L), L), {\de}^{4}G),\ \ \n(\n(L, {\de}^{2}L),
{\de}^{4}G),\ \ \n(\n(\n(\n(L, L), L), L), {\de}^{2}G),$}
\leftline{$\n(\n(\n(L, L), {\de}^2L), {\de}^{2}G),\ \ \n(\n(L, {\de}^{4}L),
{\de}^{2}G),\ \ \n(\n(\n(L, L), {\de}^{3}L), {\de}G)$}
\leftline{$\n(\n(\n(\n(\n(L, L), L), L), L), G),\ \ \n(\n(\n(\n(L, L), L),
{\de}^{2}L), G),$}
\leftline{$\n(\n(\n(L, L), {\de}^{4}L), G),\ \ \n(\n(L, {\de}^{6}L), G)$}
\leftline{$det D^{\lb1\rb}_{{23\over2}} =
{16543163447903718821855232000000\over217118368393969}(-1 + c)^{4}c^{16}(11
+ c)^{6}(21 + 4c)^{12}$}
\leftline{$\detsp (115 + 4c)(114 + 5c)(135 + 8c)^{3}(-7 + 10c)^{12}(11 +
14c)^{6}(95 + 22c)$}
\mn
 \leftline{$d$ = 12  : 23 fields :}
\leftline{$\n(\n(\n(G, {\de}G), {\de}^{2}G), {\de}^{3}G),\ \ \n(G,
{\de}^{9}G),\ \ \n(\n(L, G), {\de}^{7}G),$}
\leftline{$\n(\n(L, {\de}G), {\de}^{6}G),\ \ \n(\n(L, {\de}^{2}G),
{\de}^{5}G),\ \ \n(\n(L, {\de}^{3}G), {\de}^{4}G),$}
\leftline{$\n(\n(\n(L, L), G), {\de}^{5}G),\ \ \n(\n(\n(L, L), {\de}G),
{\de}^{4}G),\ \ \n(\n(\n(L, L), {\de}^{2}G), {\de}^{3}G),$}
\leftline{$\n(\n(\n(\n(L, L), L), G), {\de}^{3}G),\ \ \n(\n(\n(\n(L, L),
L), {\de}G), {\de}^{2}G),$}
\leftline{$\n(\n(\n(L, {\de}^{2}L), G), {\de}^{3}G),\ \ \n(\n(\n(L,
{\de}^{2}L), {\de}G), {\de}^{2}G),$}
\leftline{$\n(\n(\n(\n(\n(L, L), L), G), {\de}G), L),\ \ \n(\n(\n(\n(L,
{\de}^{2}L), G), {\de}G), L) ,$}
\leftline{$\n(\n(\n(L, {\de}^{4}L), G), {\de}G),\ \ \n(\n(\n(\n(\n(L, L),
L), L), L), L),$}
\leftline{$\n(\n(\n(\n(L, L), L), L), {\de}^{2}L),\ \ \n(\n(\n(L, L), L),
{\de}^{4}L),$}
\leftline{$\n(\n(\n(L, L), {\de}^{2}L), {\de}^{2}L),\ \ \n(\n(L, L),
{\de}^{6}L),\ \ \n(\n(L, {\de}^{2}L), {\de}^{4}L),\ \ \n(L, {\de}^{8}L)$}
\leftline{$det D^{\lb1\rb}_{12}   =
{69893115257789204783084854925269516104499200000000000000000
\over20710629462273412554478537721}(-1 + c)^{6}c^{23}$}
\leftline{$\detsp (11 + c)^{8}(21 + 4c)^{16}(115 + 4c)(114 + 5c)^{2}(135 +
8c)^{4}(-7 + 10c)^{17}$}
\leftline{$\detsp (11 + 14c)^{9}(95 + 22c)^{2}(161 + 26c)(-81 + 70c)$}
\mn
\leftline{$d ={25\over2}$  : 19 fields :}
\leftline{$\n(\n(G, {\de}G), {\de}^{7}G),\ \ \n(\n(\n(L, G), {\de}G),
{\de}^{5}G),$}
\leftline{$\n(\n(\n(L, G), {\de}^{2}G), {\de}^{4}G),\ \ \n(\n(\n(L,
{\de}G), {\de}^{2}G), {\de}^{3}G),$}
\leftline{$\n(L, {\de}^{9}G),\ \ \n(\n(\n(\n(L, L), G), {\de}G),
{\de}^{3}G),\ \ \n(\n(L, L), {\de}^{7}G),$}
\leftline{$\n(\n(\n(L, L), L), {\de}^{5}G),\ \ \n(\n(L, {\de}^{2}L),
{\de}^{5}G),\ \ \n(\n(\n(\n(L, L), L), L), {\de}^{3}G),$}
\leftline{$\n(\n(\n(L, L), {\de}^{2}L), {\de}^{3}G),\ \ \n(\n(L,
{\de}^{4}L), {\de}^{3}G),$}
\leftline{$\n(\n(\n(L, L), {\de}^{3}L), {\de}^{2}G),\ \ \n(\n(\n(\n(\n(L,
L), L), L), {\de}G), L),$}
\leftline{$\n(\n(\n(\n(L, L), {\de}^{2}L), {\de}G), L),\ \ \n(\n(\n(L, L),
{\de}^{4}L), {\de}G) ,$}
\leftline{$\n(\n(L, {\de}^{6}L), {\de}G),\ \ \n(\n(\n(\n(L, L),
{\de}^{3}L), G), L),\ \ \n(\n(\n(L, L), {\de}^5L), G)$}
\leftline{$det D^{\lb1\rb}_{{25\over2}} =
-{35952348246922073178578114668020015110823318926445772800000000000000000000
\over12565355615115842815180391716848789977906099}$}
\leftline{$\detsp (-1 + c)^{5}c^{19}(11 + c)^{7}(21 + 4c)^{14}(114 +
5c)(135 + 8c)^{3}(-7 + 10c)^{15}$}
\leftline{$\detsp (11 + 14c)^{8}(95 + 22c)(161 + 26c)(-81 + 70c)$}
\mn
}
\mn\mn
\leftline{{\bf Appendix B: A basis of quasiprimary fields up to
                           dimension $d(\Phi)$+6 built}}
\leftline{{\bf \ \ \ \ \ \ \ \ \ \ \ \ \ \ \ \ \ \  up by one
                           super primary field $\Phi$ and $\cal L$ }}
{
\mn
 \leftline{$d$ = $d(\Phi)$ : 1 field :}
 \leftline{$\phi$}
\mn
 \leftline{$d$ = $d(\Phi)$+$1\over2$  : 1 field :}
 \leftline{$\psi$}
\mn
 \leftline{$d$ = $d(\Phi)$+$3\over2$  : 1 field :}
 \leftline{$\n(\phi, G)$}
\mn
 \leftline{$d$ = $d(\Phi)$+2   : 2 fields :}
\leftline{$\n(\phi, L),\ \ \n(\psi, G)$}
\mn
 \leftline{$d$ = $d(\Phi)$+$5\over2$  : 2 fields :}
\leftline{$\n(\phi, {\de}G) , \ \ \n(\psi, L)$}
\mn
 \leftline{$d$ = $d(\Phi)$+3   : 2 fields :}
\leftline{$\n(\phi, {\de}L),\ \ \n(\psi, {\de}G) $}
\mn
 \leftline{$d$ = $d(\Phi)$+$7\over2$  : 3 fields :}
\leftline{$\n(\phi, {\de}^{2}G),\ \ \n(\n(\phi, L), G),\ \ \n(\psi,
{\de}L)$}
\mn
 \leftline{$d$ = $d(\Phi)$+4  : 5 fields :}
\leftline{$\n(\n(\phi, G), {\de}G),\ \ \n(\n(\phi, L), L),\ \ \n(\phi,
{\de}^{2}L),$}
\leftline{$\n(\psi, {\de}^{2}G),\ \ \n(\n(\psi, L), G)$}
\mn
 \leftline{$d$ = $d(\Phi)$+$9\over2$  : 6 fields :}
\leftline{$\n(\phi, {\de}^{3}G),\ \ \n(\n(\phi, L), {\de}G),\ \ \n(\n(\phi,
{\de}L), G),$}
\leftline{$\n(\n(\psi, G), {\de}G),\ \ \n(\n(\psi, L), L),\ \ \n(\psi,
{\de}^{2}L)$}
\mn
 \leftline{$d$ = $d(\Phi)$+5 : 6 fields :}
\leftline{$\n(\n(\phi, G), {\de}^{2}G),\ \ \n(\n(\phi, L), {\de}L),\ \
\n(\phi, {\de}^{3}L),$}
\leftline{$\n(\psi, {\de}^{3}G),\ \ \n(\n(\psi, L), {\de}G),\ \ \n(\n(\psi,
{\de}L), G)$}
\mn
\mn
 \leftline{$d$ = $d(\Phi)$+$11\over2$ : 8 fields :}
\leftline{$\n(\phi, {\de}^{4}G),\ \ \n(\n(\phi, L), {\de}^{2}G),\ \
\n(\n(\phi, {\de}L), {\de}G),$}
\leftline{$\n(\n(\n(\phi, L), L), G),\ \ \n(\n(\phi, {\de}^{2}L), G),\ \
\n(\n(\psi, G), {\de}^{2}G),$}
\leftline{$\n(\n(\psi, L), {\de}L),\ \ \n(\psi, {\de}^{3}L)$}
\mn
\mn
 \leftline{$d$ = $d(\Phi)$+6  : 12 fields :}
\leftline{$\n(\n(\phi, G), {\de}^{3}G),\ \ \n(\n(\phi, {\de}G),
{\de}^{2}G),\ \ \n(\n(\n(\phi, L), G), {\de}G),$}
\leftline{$\n(\n(\n(\phi, L), L), L),\ \ \n(\n(\phi, L), {\de}^{2}L),\ \
\n(\n(\phi, {\de}L), {\de}L),$}
\leftline{$\n(\phi, {\de}^{4}L),\n(\psi, {\de}^{4}G),\ \ \n(\n(\psi, L),
{\de}^{2}G),$}
\leftline{$\n(\n(\psi, {\de}L), {\de}G),\ \ \n(\n(\n(\psi, L), L), G),\ \
\n(\n(\psi, {\de}^{2}L), G)$}
\mn
\mn
}
\leftline{{\bf Appendix C: Determinants of quasiprimary parts
                           of the Kac-matrix }}
\leftline{{\bf \ \ \ \ \ \ \ \ \ \ \ \ \ \ \ \ \ \
                           with one additional super primary field
                           involved }}
{
\mn
\leftline{$\sw$($3\over2$,2) :}
\leftline{$det D^{\lb2\rb}_{{7\over2}} = {1\over15}c(6 + 5c)$}
\leftline{$det D^{\lb2\rb}_{4}   = -{1\over150}c^{2}(29 + 2c)(6 + 5c)$}
\mn
\leftline{$\sw$($3\over2$,$7\over2$) :}
\leftline{$det D^{\lb{7\over2}\rb}_{5} \  = -{1\over21}c(21 + 4c)$}
\leftline{$det D^{\lb{7\over2}\rb}_{{11\over2}} = {1\over336}c^{2}(53 +
2c)(21 + 4c)$}
\leftline{$det D^{\lb{7\over2}\rb}_{6}  \  = {1\over40}(-1 + c)c^{2}(53 +
2c)$}
\leftline{$det D^{\lb{7\over2}\rb}_{{13\over2}} = {4\over121}(-1 +
c)c^{2}(21 + 4c)$}
\mn
\leftline{$\sw$($3\over2$,4) :}
\leftline{$det D^{\lb4\rb}_{{11\over2}} = {1\over18}c(20 + 3c)$}
\leftline{$det D^{\lb4\rb}_{6}  \  = -{1\over324}c^{2}(61 + 2c)(20 + 3c)$}
\leftline{$det D^{\lb4\rb}_{{13\over2}} = -{1\over495}c^{2}(61 + 2c)(-7 +
10c)$}
\leftline{$det D^{\lb4\rb}_{7} \   = -{1\over270}c^{2}(20 + 3c)(-7 + 10c)$}
\leftline{$det D^{\lb4\rb}_{{15\over2}} = -{1\over16731}c^{3}(61 + 2c)(20 +
3c)^{2}(-65 + 44c)$}
\leftline{$det D^{\lb4\rb}_{8}  \  =  {1\over 29513484}c^{5}(5 + 2c)(61 +
2c)^{2}(20 + 3c)^{2}(-7 + 10c)(377 + 10c)(-65 + 44c)$}
\mn
\leftline{$\sw$($3\over2$,$11\over2$) :}
\leftline{$det D^{\lb{11\over2}\rb}_{7} \   = -{4\over33}c(11 + c)$}
\leftline{$det D^{\lb{11\over2}\rb}_{{15\over2}} = {1\over198}c^{2}(11 +
c)(85 + 2c)$}
\leftline{$det D^{\lb{11\over2}\rb}_{8}  \  = {1\over1092}c^{2}(85 + 2c)(5
+ 13c)$}
\leftline{$det D^{\lb{11\over1}\rb}_{{17\over2}} = {16\over2925}c^{2}(11 +
c)(5 + 13c)$}
\leftline{$det D^{\lb{11\over2}\rb}_{9} \   = -{2\over1155}c^{3}(11 +
c)^{2}(85 + 2c)(-10 + 7c)$}
\leftline{$det D^{\lb{11\over2}\rb}_{{19\over2}} = {1\over2203047}c^{5}(11
+ c)^{2}(85 + 2c)^{2}(-10 + 7c)(5 + 13c)(4717 + 1092c + 20c^2)$}
\leftline{$det D^{\lb{11\over2}\rb}_{10}   = {5\over20857419}c^{6}(11 +
c)^{2}(15 + c)(85 + 2c)^{2}(-5 + 4c)(5 + 13c)^{2}(4717 + 1092c + 20c^{2})$}
\leftline{$det D^{\lb{11\over2}\rb}_{{21\over2}} =
{819200\over19321000497}c^{6}(11 + c)^{3}(15 + c)(85 + 2c)^{2}(-5 + 4c)(-10
+ 7c)(5 + 13c)^{2}$}
\mn
\leftline{$\sw$($3\over2$,6) :}
\leftline{$det D^{\lb6\rb}_{{15\over2}} = {1\over117}c(162 + 13c)$}
\leftline{$det D^{\lb6\rb}_{8} \   = -{1\over3042}c^{2}(93 + 2c)(162 +
13c)$}
\leftline{$det D^{\lb6\rb}_{{17\over2}} = -{1\over1365}c^{2}(93 + 2c)(11 +
14c)$}
\leftline{$det D^{\lb6\rb}_{9}  \  = -{1\over2912}c^{2}(162 + 13c)(11 +
14c)$}
\leftline{$det D^{\lb6\rb}_{{19\over2}} = -{1\over338130}c^{3}(93 + 2c)(162
+ 13c)^{2}(-27 + 20c)$}
\leftline{$det D^{\lb6\rb}_{10}   = {1\over1602229005}c^{5}(93 +
2c)^{2}(162 + 13c)^{2}(11 + 14c)(-27 + 20c)(5917 + 1188c + 20c^{2})$}
\leftline{$det D^{\lb6\rb}_{{21\over2}} = -{16\over186752880639}c^{6}(33 +
2c)(93 + 2c)^{2}(162 + 13c)^{2}(11 + 14c)^{2}$}
\leftline{$\detsp (-27 + 20c)(5917 + 1188c + 20c^{2})$}
\leftline{$det D^{\lb6\rb}_{11}   = -{2\over4461283125}c^{6}(33 + 2c)(93 +
2c)^{2}(162 + 13c)^{3}(11 + 14c)^{2}(-27 + 20c)^{2}$}
\leftline{$det D^{\lb6\rb}_{{23\over2}} =
-{2048\over86728774951175175}c^{8}(93 + 2c)^{3}(162 + 13c)^{5}(11 +
14c)^{2}$}
\leftline{$\detsp (-27 + 20c)^{2}(-43 + 85c)(5917 + 1188c + 20c^{2}))   $}
\leftline{$det D^{\lb6\rb}_{12}  =
-{8192\over412159655278351185230025}c^{12}(33 + 2c)(93 + 2c)^{5}(162 +
13c)^{6}(11 + 14c)^{3}$}
\leftline{$\detsp (-27 + 20c)^{3}(-43 + 85c)(5917 + 1188c + 20c^{2} )^{2}
(-837 + 8540c + 140c^{2})$}
}
\mn
\mn
\leftline{{\bf Appendix D: Realization of $\bf\SW({3\over2},{11\over2})$
at $\bf c = {10\over7}$  as a subalgebra of $\bf\SW({3\over2},{5\over2})$
}}
\mn
\doublespace
As mentioned in chapter four and five there is a strange connection between
these algebras at $c = {10\over7}$. $\sw$(${3\over2},{5\over2}$)
contains a super primary field  of dimension ${11\over2}$.
The two components of  this field  are:
$$ \eqalign{
   \tilde{\phi} &= \alpha \  \bigl(  \ \  \n(\psi,\phi)  +
                         {\textstyle {8869\over29580}}
C^{\psi}_{G\phi}\n(L,{\de^2} G) -
                         {\textstyle {5537\over25143}}
C^{\psi}_{G\phi}\n(\n(L,L),G)
                     \ \     \bigr)  \cr
\tilde{\psi} &= {\alpha\over\sqrt{2}} \ \bigl( \ \ \n(\psi,\psi) +
                {\textstyle {1\over6}} \n(\phi,\de\phi) +
                {\textstyle {8575\over67048}} \n(G,{\de^3}G) + \cr
&\ \ \ \ \ \ \ \ \ \ \ {\textstyle{206143\over251430}} \n(L,{\de^2}L)  -
                       {\textstyle{5537\over25143}} \n(\n(L,G),\de G )) -
                       {\textstyle{11074\over25143}} \n(\n(L,L),L)
                       \ \     \bigr)   \cr
 \alpha^2 &= {\textstyle {176001\over1567918}}
}$$
where $\Phi=\phi +\theta\psi$ is the simple super primary field of
 dimension ${5\over2}$.
The fields  $\tilde\phi$ ,$\tilde\psi$ are orthogonal to all other
fields  of dimension ${11\over2},6$ and their normalization is
described by (3.3'). We have calculated the self-coupling:
$$ { \bigl( C^{\tilde\psi}_{\tilde{\phi}\tilde{\phi}} \bigr) }^2 = {
844918800\over34554863} $$
This is exactly the coupling obtained in the direct construction of
$\sw$(${3\over2},{11\over2}$) at $c = {10\over7}$.
The two different signs of the
coupling are realized by the two possible signs for $\alpha$.
\mn
\eject
\leftline{{\bf References}}
\bigskip
\settabs\+ & \phantom{---------} &
\phantom{-------------------------------------
-----------------------------------------}
& \cr
\+ & $\q{\bpz}$ & A.A. Belavin, A.M. Polyakov, A.B. Zamolodchikov, Nucl.
Phys. {\bf B}241 (1984) p. 333  & \cr
\+ & $\q{\zam}$ & A.B. Zamolodchikov, Theor. Math. Phys. 65 (1986) p. 1205
& \cr
\+ & $\q{\bai}$ & F.A. Bais, P. Bouwknegt, M. Surridge, K. Schoutens, Nucl.
Phys. {\bf B}304 (1988) p. 371 & \cr
\+ & $\q{\blg}$ & A. Bilal, J.L. Gervais, Nucl. Phys. {\bf B}318 (1989) p.
579 & \cr
\+ & $\q{\bal}$ & J. Balog, L. Feh\'er, P. Forg\'acs, L. O'Raifeartaigh, A.
Wipf & \cr
\+ &            & Phys. Lett. {\bf B}244 (1990) p. 435 & \cr
\+ & $\q{\ham}$ & K. Hamada, M. Takao, Phys. Lett. {\bf B}209 (1988) p. 247
& \cr
\+ &            & Erratum, Phys. Lett. {\bf B}213 (1988) p. 564 & \cr
\+ & $\q{\zha}$ & D.H. Zhang, Phys. Lett. {\bf B}232 (1989) p. 323 & \cr
\+ & $\q{\fig}$ & J.M. Figueroa-O'Farrill, S. Schrans, Phys. Lett. {\bf
B}245 (1990) p. 471 & \cr
\+ & $\q{\bou}$ & P. Bouwknegt, Phys. Lett. {\bf B}207 (1988) p. 295 & \cr
\+ & $\q{\blm}$ & R. Blumenhagen, M. Flohr, A. Kliem, W. Nahm, A.
Recknagel, R. Varnhagen & \cr
\+ &            & Nucl. Phys. {\bf B}361 (1991) p. 255 & \cr
\+ & $\q{\kau}$ & H.G. Kausch, G.M.T. Watts,  Nucl. Phys. {\bf B}354 (1991)
p. 740 & \cr
\+ & $\q{\wer}$ & W. Nahm, in proceedings Trieste Conference on & \cr
\+ &            & 'Recent Developments in Conformal Field Theories', ICTP,
Trieste (1989) p. 81 & \cr
\+ & $\q{\fri}$ & D. Friedan, Z. Qiu, S. Shenker, Phys. Lett. {\bf B}151
(1985) p. 37& \cr
\+ & $\q{\ina}$ & T. Inami, Y. Matsuo, I. Yamanaka,  Phys. Lett. {\bf B}215
(1988) p. 701 & \cr
\+ & $\q{\kom}$ & S. Komata, K. Mohri, H. Nohara,  Nucl.  Phys. {\bf B}359
(1991) p. 168 & \cr
\+ & $\q{\ho1}$ & K. Hornfeck, E. Ragoucy, Nucl. Phys. {\bf B}340 (1990) p.
225 & \cr
\+ & $\q{\jos}$ & J.M. Figueroa-O'Farrill, S. Schrans, Int. Jour. of Mod.
Phys. {\bf A}7 (1992) p. 591 & \cr
\+ & $\q{\blu}$ & R. Blumenhagen, preprint BONN-HE-91-20 to be published in
Nucl. Phys. {\bf B} & \cr
\+ & $\q{\hor}$ & K. Hornfeck, preprint London, King's College 91.08.26 &
\cr
\+ & $\q{\nam}$ & W. Nahm, {\it Conformal Quantum Field Theories in Two
Dimensions} & \cr
\+ &            & World Scientific, to be published & \cr
\+ & $\q{\wes}$ & P. West, {\it Introduction to Supersymmetry and
Supergravity } & \cr
\+ &            & World Scientific (1986) & \cr
\+ & $\q{\flo}$ & M. Flohr, Diplomarbeit BONN-IR-91-30 & \cr
\+ & $\q{\bo1}$ & P. Bouwknegt, in proceedings of the meeting 'Infinite
dimensional Lie algebras and Groups'& \cr
\+ &            & C.I.R.M., Luminy, Marseille (1988) p. 527& \cr
\+ & $\q{\cap}$ & A. Cappelli, C. Itzykson, J.B. Zuber, Comm. Math. Phys.
113 (1987) p. 1 & \cr
\+ & $\q{\ka1}$ & H.G. Kausch, Phys. Lett. {\bf B}259 (1991) p. 448 & \cr
\+ & $\q{\mfl}$ & M. Flohr, preprint BONN-HE-92-08 & \cr
\+ & $\q{\eho}$ & W. Eholzer, M. Flohr , A. Honecker , R. H{\"u}bel, W.
Nahm, R. Varnhagen  & \cr
\+ &            & preprint BONN-HE-91-22 to be published in Nucl. Phys.
{\bf B} & \cr
\+ & $\q{\ber}$ & M.A. Bershadsky, V.G. Knizhnik, M.G. Teitelman, Phys.
Lett. {\bf B}151 (1985) p. 31 & \cr
\+ & $\q{\eic}$ & H. Eichenherr, Phys. Lett. {\bf B}151 (1985) p. 26 & \cr
\+ & $\q{\mus}$ & G. Mussardo, G. Sotkov, M. Stanishkov, Phys. Lett. {\bf
B}195 (1987) p. 397, & \cr
\+ &            & Nucl. Phys. {\bf B}305 (1988) p.69 & \cr
\+ & $\q{\god}$ & P. Goddard, A. Kent, D. Olive,  Comm. Math. Phys. 103
(1986) p. 105 & \cr
\+ & $\q{\sch}$ & A.N. Schellekens, S. Yankielowicz, Nucl. Phys. {\bf B}327
(1989) p. 673 & \cr
\+ & $\q{\dop}$ & S. Doplicher, R. Haag, J.E. Roberts, Comm. Math. Phys. 23
(1971) p. 199,& \cr
\+ &            & Comm. Math. Phys. 35 (1974) p. 49  & \cr
\+ & $\q{\fre}$ & K. Fredenhagen, K.H. Rehren, B. Schroer, Comm. Math.
Phys. 125 (1989) p. 201 & \cr
\+ & $\q{\wil}$ & F. Wilczek, Phys. Rev. Lett. 49 (1982) p. 957  & \cr
\+ & $\q{\eh1}$ & W. Eholzer, A. Honecker, R. H{\"u}bel & \cr
\+ &            & {\it Representations of N = 1 Extended Superconformal
Algebras  }, in preparation  & \cr
\+ & $\q{\del}$ & F. Delduc, E. Ragoucy, P. Sorba, preprint Lyon,
ENSLAPP-L-352-91 & \cr
\+ & $\q{\sou}$ & K. Schoutens, A. Sevrin, Phys. Lett. {\bf B}258 (1991) p.
134 & \cr
\+ & $\q{\rva}$ & R. Varnhagen, Phys. Lett. {\bf B}275 (1992) p. 87 & \cr
\vfill
\end